\documentclass[letterpaper]{emulateapj}
\usepackage{wasysym} 
\usepackage{graphicx}
\usepackage{subfigure}
\usepackage{natbib}

\usepackage{amsmath}

\newcommand{\cii}{C\,{\sc ii}}
\newcommand{\hii}{H\,{\sc ii}}
\newcommand{\mum}{\ensuremath{\mu \rm{m}}}

\begin{document}

\title{Variations of the ISM Compactness Across the Main Sequence of Star Forming Galaxies: Observations and Simulations} 

\shorttitle{ISM compactness across the MS}

\author{J.~R. Mart\'{\i}nez-Galarza\altaffilmark{1}, H.~A. Smith\altaffilmark{1}, L. Lanz \altaffilmark{2}, Christopher C. Hayward\altaffilmark{3,1}, A. Zezas\altaffilmark{4,1}, L. Rosenthal\altaffilmark{5}, A. Weiner\altaffilmark{1,6}, C. Hung\altaffilmark{1,7}, M.~L.~N. Ashby\altaffilmark{1}, B. Groves\altaffilmark{8}}
\altaffiltext{1}{Harvard-Smithsonian Center for Astrophysics, 60 Garden Street, Cambridge, MA 02138, USA, jmartine@cfa.harvard.edu}
\altaffiltext{2}{Infrared Processing and Archival Center, California Institute of Technology, MC 100-22, Pasadena, CA 91125, USA}
\altaffiltext{3}{TAPIR 350-17, California Institute of Technology, 1200 E. California Blvd., Pasadena, CA 91125, USA}
\altaffiltext{4}{University of Crete, Physics Department, P.O. Box 2208, 710 03 Heraklion, Crete, Greece}
\altaffiltext{5}{Haverford College, 370 Lancaster Ave, Haverford, PA 19041, USA}
\altaffiltext{6}{RPI Institute, 110 8th St, Troy, NY 12180, USA}
\altaffiltext{7}{Institute for Astronomy, 2680 Woodlawn Drive, Honolulu, HI 96822-1839, USA}
\altaffiltext{8}{Max-Planck-Institut f\"ur Astronomie, K\"onigstuhl 17, 69117, Heidelberg, Germany}

\begin{abstract}

The majority of star-forming galaxies follow a simple empirical correlation in the star formation rate (SFR) versus stellar mass ($M_*$) plane, of the form $\mbox{SFR} \propto M_*^{\alpha}$, usually referred to as the star formation Main Sequence (MS). The physics that sets the properties of the MS is currently a subject of debate, and no consensus has been reached regarding the fundamental difference between members of the sequence and its outliers. Here we combine a set of hydro-dynamical simulations of interacting galactic disks with state-of-the-art radiative transfer codes to analyze how the evolution of mergers is reflected upon the properties of the MS. We present \textsc{Chiburst}, a Markov Chain Monte Carlo (MCMC) Spectral Energy Distribution (SED) code that fits the multi-wavelength, broad-band photometry of galaxies and derives stellar masses, star formation rates, and geometrical properties of the dust distribution. We apply this tool to the SEDs of simulated mergers and compare the derived results with the reference output from the simulations. Our results indicate that changes in the SEDs of mergers as they approach coalescence and depart from the MS are related to an evolution of dust geometry in scales larger than a few hundred parsecs. This is reflected in a correlation between the specific star formation rate (sSFR), and the compactness parameter $\mathcal{C}$, that parametrizes this geometry and hence the evolution of dust temperature ($T_{\rm{dust}}$) with time. As mergers approach coalescence, they depart from the MS and increase their compactness, which implies that moderate outliers of the MS are consistent with late-type mergers. By further applying our method to real observations of Luminous Infrared Galaxies (LIRGs), we show that the merger scenario is unable to explain these extreme outliers of the MS. Only by significantly increasing the gas fraction in the simulations are we able to reproduce the SEDs of LIRGs.

\end{abstract}

\section{Introduction}
\label{sec:intro}

Over the last decade, deep optical and infrared surveys have established that the majority of star-forming galaxies up to $z=2.5$ follow a simple scaling correlation linking their stellar mass with their star formation rate (SFR): $\mbox{SFR} = M_*^{\alpha}$, with $0.5 < \alpha < 1.0$ \citep{Brinchmann04, Noeske07, Elbaz07, Daddi07, Rodighiero11}. The bulk of star-forming galaxies that lie near this simple correlation have been collectively called the ``main sequence'' (MS) of star-forming galaxies. The existence of a MS has been interpreted as evidence that the majority of galaxies throughout cosmic history form stars in a steady, secular mode, in time scales that are longer than their dynamical timescales \citep{Genzel10, Wuyts11}, a mode that requires a continuous replenishment of gas from the intergalactic medium. In this picture, the outliers above the MS would be explained by starburst events, probably triggered by mergers, with shorter depletion timescales yielding larger specific star formation rates (sSFRs) for systems off the MS. It has been noted that the zero-point of the MS evolves with redshift, with the MS in the intermediate- and high-$z$ Universe located higher in SFR with respect to the local MS. A possible interpretation of this behavior is given in terms of a larger gas fraction ($f_{\rm{gas}}$) in galaxies earlier in cosmic history \citep[e.g.][]{Scoville15}.

Both simulations \citep[e.g.][]{Dave11,Torrey14,Sparre14} and semi-analytical models \citep[e.g.][]{Dutton10,Dave12,Dekel13,Mitchell14} predict the existence of the MS and qualitatively reproduce the redshift dependence of the normalization. In these models, the redshift evolution is driven by evolution in the gas accretion rates onto galaxies. The scatter in the MS at fixed stellar mass, which is $\sim0.2-0.4$ dex, may be caused by variations in gas accretion rates, formation histories, environment, or galaxy structure, among other possible causes. Merger-induced starbursts can cause galaxies to temporarily move significantly above the MS, but state-of-the-art cosmological hydrodynamical simulations underpredict the abundance of MS outliers \citep{Sparre14}. This underprediction may suggest that other processes, such as violent disk instability \citep[e.g.][]{Ceverino10, Porter14} may be an important source of outliers above the MS. It has also been suggested that the scatter of the MS may simply be a consequence of the central limit theorem if \emph{in situ} star formation is a stochastic process \citep{Kelson14}. The approximately linear slope and the redshift evolution of the MS may therefore not be a useful constraint on theoretical models. On the other hand, the magnitude of the MS scatter, as well as the properties of the population of MS outliers may be able to provide more useful constraints on the physics controlling star formation. It is therefore of crucial importance to quantify the variations in physical properties of galaxies across the MS.

Despite recent observational and theoretical progress, a convincing picture explaining why certain galaxies belong to the MS whereas other are outliers of this scaling relation, or why the normalization of the MS evolves with redshift (implying that galaxies at earlier cosmic times had on average larger sSFRs), remains elusive. In particular, it is not clear whether the latter is a consequence of larger gas reservoirs in the early Universe, or if galaxies were more efficient at converting gas into stars early on. Recent \textit{Atacama Large Millimeter Array} (ALMA) observations suggests that all galaxies at all cosmic times convert gas into stars with the same efficiency, and that the higher sSFRs observed at intermediate- and high-$z$ are due to relatively larger gas reservoirs \citep{Scoville15}. On the other hand, using a multi-wavelength stacking analysis of the spectral energy distribution (SEDs) of galaxies in the COSMOS field, \citet{Bethermin15} argue that, although the bulk of star formation up to $z\sim 4$ is dominated by secular processes, strong starburst and MS galaxies have different star formation efficiencies. Studying how star formation properties such as dust geometry and star formation rates evolve across the MS, and whether they are different on and off the sequence can provide additional evidence to settle these open issues. 

It is well established that the majority of galaxies that constitute the MS have disk morphologies, although disturbed morphologies can also be found within the sequence. On the other hand, while the majority of outliers appear to have cuspier morphologies that indicate late stages of mergers \citep{Hung13, Wuyts11}, disk morphologies are also found outside the MS. The role of mergers in shaping the properties of the MS is therefore not straightforward. Hydrodynamical simulations of isolated and interacting systems can be used in combination with observations of real systems at different stages of interactions in order to evaluate membership to the MS as a function of stellar mass, interaction stage, and gas content. In this paper, we use a statistically robust SED analysis method (\textsc{Chiburst}), which is based on the star-forming galaxy models of \citet{Groves08}, to consistently analyze the star-forming properties of a set of hydro-dynamical simulations of binary mergers and compare them with real observations of local mergers, local luminous ($L>10^{11}\, \mbox{L}_{\odot}$) mergers, and intermediate-$z$ Luminous Infrared Galaxies (LIRGs). The goal is to investigate how mergers affect the membership of individual galaxies to the MS as a function of merger stage, and how mobility across the sequence relates to the geometry of dust distribution within these systems.  

Our methodology is straightforward: we first apply \textsc{Chiburst} to the multi-wavelength mock SEDs of a set of simulated binary mergers with different mass ratios. These SEDs are calculated at different stages between first approach and coalescence by performing dust radiative transfer in post-processing on hydrodynamical simulations with an injected empirical star formation law. We use the results to study how the evolution of a merger affects the location of a particular galaxy with respect to the MS, and how this affects the large scale geometry of the dust (i.e., its compactness) and therefore the distribution of dust temperatures. To put real galaxies in context, we then apply the same SED fitting method to a sample of 24 local interacting galaxies at different stages, 6 local luminous ($L>10^{11}\, \mbox{L}_{\odot}$) late-type mergers, and 9 $z\sim 0.3$ LIRGs for which ultraviolet to far-infrared (far-IR) photometry is available. A direct comparison of the results between simulated and real systems allows us to infer whether mergers are generally the best explanation for the population of MS outliers.

The paper is structured as follows. In \S~\ref{sec:observations} we describe the observations of interacting galaxies used in this paper as well as the reduction process used to obtain their SEDs. In \S~\ref{sec:hydro_models} we give an overview of the hydrodynamical models and the radiative transfer code used to generate the mock SEDs of interacting systems. We present \textsc{Chiburst}, our novel Bayesian Monte Carlo fitting method in \S~\ref{sec:fitting}, and in \S~\ref{sec:results} we show the results of applying it to both the observed galaxies and the simulated ones. We discuss the correlations found and their implications for the nature of the MS in \S~\ref{sec:discuss}. Finally, we summarize our findings in \S~\ref{sec:summary}.

\section{Observations}
\label{sec:observations}
The sample of observed galaxies analyzed here comprises three different groups: \textit{(i):} a subsample of 27 local interactions from the \emph{Spitzer} Interactive Galaxy Survey (SIGS) \citep{Brassington15}; \textit{(ii):} an additional group of 6 local, luminous ($>10^{11}~\mbox{L}_{\odot}$) stage 4 interactions that we have morphologically selected from the Sloan Digital Sky Survey (SDSS) and Galaxy Zoo \citep{Lintott08} and cross-checked with the revised \emph{IRAS} Faint Source Catalog \citep{Moshir92}; and \textit{(iii): }a group of 9 \emph{Herschel}-selected intermediate-redshift ($z\sim 3$) LIRGs from \citet{Magdis14}. We list the observed systems in Table~\ref{tab:obs_list}.

\begin{deluxetable*}{ccccc}
\tablecolumns{9}
\tablecaption{Observed galaxies}

\tablehead{
  \colhead{Galaxy} &
  \colhead{Type} &
  \colhead{$\alpha$} &
  \colhead{$\delta$} &
  \colhead{$z$} 
}
\startdata

 NGC~2976 & Local merger  & 09 47 16.3  & +67 54 52.0  & 0.0009 \\ 
 NGC~3031 & Local merger  & 09 55 33.2  & +69 03 57.9  & 0.0009 \\
 NGC~3034 & Local merger  & 09 55 52.2  & +69 40 47.8  & 0.0009 \\
 NGC~3185 & Local merger  & 10 17 38.7  & +21 41 16.2  & 0.0053 \\
 NGC~3187 & Local merger  & 10 17 48.4  & +21 52 30.9  & 0.0061 \\
 NGC~3190 & Local merger  & 10 18 05.7  & +21 49 57.0  & 0.0053 \\
 NGC~3395/3396 & Local merger  & 10 49 50.0  & +32 58 55.2  & 0.0065 \\
 NGC~3424 & Local merger  & 10 51 46.9  & +32 54 04.1  & 0.0061 \\
 NGC~3430 & Local merger  & 10 52 11.5  & +32 57 05.0  & 0.0062 \\
 NGC~3448 & Local merger  & 10 54 38.7  & +54 18 21.0  & 0.0057 \\
 UGC~6016 & Local merger  & 10 54 13.4  & +54 17 15.5  & 0.0064 \\ 
 NGC~3690/IC694 & Local merger  & 11 28 31.2  & +58 33 46.7  & 0.0112 \\
 NGC~3786 & Local merger  & 11 39 42.5  & +31 54 34.2 & 0.0097 \\
 NGC~3788 & Local merger  & 11 39 44.6  & +31 55 54.3  & 0.0085 \\
 NGC~4038/4039 & Local merger  & 12 01 53.9  & −18 52 34.8  & 0.0062 \\
 NGC~4618 & Local merger  & 12 41 32.8  & +41 08 44.4  & 0.0017 \\
 NGC~4625 & Local merger  & 12 41 52.6  & +41 16 20.6  & 0.0019 \\
 NGC~4647 & Local merger  & 12 43 32.6  & +11 34 53.9  & 0.0039 \\
 M~51a  & Local merger  & 13 29 54.1  & +47 11 41.2  &   0.0018  \\
 M~51b  & Local merger  & 13 29 59.7  & +47 15 58.5 & 0.0018   \\
 NGC~5394 & Local merger  & 13 58 33.7  & +37 27 14.4  & 0.0131 \\
 NGC~5395 & Local merger  & 13 58 37.6  & +37 25 41.2  & 0.0131 \\
 M~101  & Local merger  & 14 03 09.8  & +54 20 37.3  &   0.0015 \\
 NGC~5474 & Local merger  & 14 05 01.2  & +53 39 11.6  & 0.0014 \\
 \hline
 NGC~2623  & Local LIRG  & 08 38 24.1  & +25 45 16.7 & 0.0185 \\
 UGC~4881  & Local LIRG  & 09 15 55.5  & +44 19 58.2 &  0.0392 \\
 VV~283  & Local LIRG  & 13 01 50.3  & +04 20 00.5 & 0.0374 \\
 Mrk~273  & Local LIRG  & 13 44 42.1  & +55 53 13.2 & 0.0373 \\ 
 VV~705  & Local LIRG  & 15 18 06.1  & +42 44 44.6 & 0.0400 \\
 NGC~6090  & Local LIRG  & 16 11 40.4  & +52 27 21.5 &  0.0294 \\ 
 \hline
 ELAISS  & Interm. $z$ LIRG  & 00 40 14.6  & −43 20 10.1 & 0.265 \\
 CDFS2  & Interm. $z$ LIRG  & 03 28 18.0   & −27 43 08.0 & 0.248 \\ 
 CDFS1  & Interm. $z$ LIRG  & 03 29 04.3   & −28 47 52.9 & 0.289 \\ 
 SWIRE4  & Interm. $z$ LIRG  & 10 32: 37.4  & +58 08 46.0 & 0.251 \\
 SWIRE5  & Interm. $z$ LIRG  & 10 35 57.9  & +58 58 46.2 & 0.366 \\ 
 SWIRE2  & Interm. $z$ LIRG  & 10 51 13.4  & +57 14 26.2 & 0.362 \\
 SWIRE7  & Interm. $z$ LIRG  & 11 02 05.7  & +57 57 40.6 & 0.550 \\
 BOOTES2  & Interm. $z$ LIRG  & 14 32 34.9  & +33:28:32.3 & 0.250 \\
 BOOTES1  & Interm. $z$ LIRG  & 14 36 31.9  & +34 38 29.1 & 0.354

\enddata

\label{tab:obs_list}
\end{deluxetable*}

The SIGS sample is fully described in \citet{Brassington15} and was designed to span a broad range of galaxy interaction properties based on their interaction probability and not only on morphological properties. This was done to guarantee that the sample included systems covering the full range of interaction stages and not only those with strong morphological disturbances. In this paper we analyze the subsample presented in \citet[][L13 hereafter]{Lanz13}, which include those SIGS galaxies for which \emph{Herschel} data was available when that paper was prepared. Out of the 31 galaxies in L13, we exclude NGC~3226, NGC~3227, and NGC~3077, because they do not have reliable \textit{GALEX} data. We also exclude NGC~4649, a large elliptical with very little mid-infrared/far-infrared emission. Additionally, the pairs NGC~3395/96, NGC~3690/IC~694, and NGC~4038/4039 are indistinguishable within a single aperture, which means that we only have their integrated SEDs. We therefore have a total of 24 SEDs of local interactions. In L13, these galaxies have been ranked by interactions stage between stage 1 (isolated, non-interacting galaxies) and stage 4 (strongly interacting galaxies) and all except one (NGC~3690) have luminosities between $0.25\times 10^9~\mbox{L}_{\odot}$ and $9.8\times 10^{10}~\mbox{L}_{\odot}$. From their \emph{Spitzer} mid-infrared (mid-IR) colors, none of them appears to be globally dominated by active galactic nuclei (AGN) activity. 

The photometry for these galaxies is fully described in L13. Briefly, matched apertures were used across all wavebands, choosing in each case the Kron aperture needed to fully encircle the galaxy in the waveband where it looks most extended. A local background was applied in each case, and aperture corrections were applied.

The galaxies in the second group were chosen to extend our sample of local interactions to include bright, late-type interactions that had available \emph{Herschel} observations. They were selected by cross-referencing the \emph{IRAS} Faint Source Catalog with Galaxy Zoo objects that show stage 4 morphology and have luminosities $L>10^{11}~\rm{L}_{\odot}$. All six objects selected this way had \emph{Herschel}-PACS and \emph{Herschel}-SPIRE maps available from the Great Observatories All-sky LIRG Survey (GOALS); \citep{Armus09}. We reduced these maps using the \emph{Herschel} Interactive Processing Environment (HIPE); \citep{Ott10} and performing aperture photometry for our galaxies using the same method as for the L13 galaxies.  

Finally, the intermediate-$z$ systems were selected from the original \emph{Herschel} Multi-tiered Extragalactic Survey \citep[HERMES;][]{Oliver12}. Only sources with $S_{250} > 150\, \mbox{mJy}$ in fields covered by the survey and with available spectrometric redshifts were selected. The redshift range covered is $0.248< z <0.366$, except for one object whose redshift is 0.550. The luminosities of these systems are similar to those in the second group of local interactions, although they are all spatially unresolved, and therefore we can not confirm that they are merger systems. The full sample and the photometry extraction is described in \citet{Magdis14}.
 
\section{Simulations and mock photometry of interacting systems}
\label{sec:hydro_models}
\subsection{Hydrodynamics}
We use the same suit of hydrodynamical simulations described in \citet[][L14 hereafter]{Lanz14}, with some additions. Here we will briefly describe those aspects of the simulations that are relevant for our discussion. A hydrodynamical code is combined with a radiative transfer code to obtain mock SEDs at different times for four isolated progenitor galaxies (M0,M1,M2,M3) and the 10 possible binary mergers arising from interactions between these progenitors. These four progenitor galaxies are similar to typical SDSS galaxies, with masses ranging from $6 \times 10^8\, \mbox{M}_{\odot}$ to $4 \times 10^{10}\, \mbox{M}_{\odot}$. Here we add a fifth progenitor galaxy (M4) with a starting stellar mass of $1.25\times 10^{11}\, M_{\odot}$, which is three times more massive than M3. This progenitor is intended to be an analog of a massive spiral that has consumed most of its initial gas mass by forming stars. Its properties are those of the late-type spirals observed in the local Universe. For this M4 galaxy, we only simulate a binary interaction with itself. We also include a more massive, gas-rich merger intended to represent a typical submillimeter galaxy (SMG) without a well-defined bulge. Such galaxies are exclusively found at high redshifts ($z>1$). We list the basic properties for our simulated mergers in Table \ref{tab:progenitors}.

The hydrodynamical calculations are performed using the \textsc{Gadget-3} code \citep{Springel05}, which solves the gas dynamics using smoothed particle hydrodynamics (SPH)\footnote{The simulations used the entropy-conserving version of SPH. However, we do not believe that this is a significant limitation, because \citet{Hayward14a} demonstrated that for such idealized simulations that employ an effective equation of state treatment of the ISM, the results yielded by the state-of-the-art moving-mesh code \textsc{Arepo} \citep{Springel10} are very similar to those yielded by the standard SPH formulation in \textsc{Gadget}-3, which was used here}. The radiative transfer is calculated using the 3D Monte Carlo dust radiative transfer code \textsc{Sunrise} \citep{Jonsson06, Jonsson10}. The methods for combining the output from the hydrodynamical code with the dust physics for particular types of galaxies are described in \citet{Narayanan10a, Narayanan10b, Hayward11} and \citet{Hayward12}. 

\begin{deluxetable}{ccccccc}
\tablecolumns{7}
\tablecaption{Galaxy Models for the simulations}

\tablehead{
  \colhead{} &
  \colhead{M0} &
  \colhead{M1} &
  \colhead{M2} &
  \colhead{M3} &
  \colhead{M4} &
  \colhead{SMG} 
}
\startdata

$M_*$ ($10^{10}~\mbox{M}_{\odot}$) & 0.061  &  0.38  &  1.128  &  4.22 & 12.5 & 16.0 \\
$M_{\mbox{tot}}$ ($10^{10}~\mbox{M}_{\odot}$) &  5.0  &  20.0  &  51.0  & 116.0 & 260.0 & 940.0 \\
$M_{\rm{gas}}$ ($10^{10}~\mbox{M}_{\odot}$) &  0.035  &  0.14  &  0.33  &  0.80 & 1.8  & 24.0 \\
$N_{\mbox{DM}}$ &  30000  &  50000  &  80000  &  12000 & 26400 & 60000 \\
$N_{\rm{gas}}$  &  10000  &  20000  &  30000  &  50000 & 32000 & 48000 \\

\enddata

\label{tab:progenitors}
\end{deluxetable}

Unresolved star formation is accounted for by assuming that gas particles with densities above $n\sim 0.1~\mbox{cm}^{-3}$ form stars according to the volume-density-dependent Schmidt-Kennicutt (S-K) law $\dot{\rho_*} \propto n_{\rm{gas}}^{\alpha}$, with $\alpha = 1.5$ \citep{Schmidt59, Kennicutt98}. Due to the limited mass resolution of the simulations, individual stars are not created. Instead, equal mass star particles are generated stochastically such that the SFR obtained in this way agrees with the S-K law. For all snapshots in each simulation, we register the following quantities: the instantaneous SFR ($\mbox{SFR}_{\mbox{inst}}$), defined as the sum of the SFRs of the individual gas particles, calculated based on their gas densities and the assumed sub-resolution star formation prescription; the separation between the nuclei of the interacting pair ($d_{\rm{BH}}$); and the mass of all stars formed from the start of the simulation ($m_{\rm{N}}$).

\subsection{Radiative transfer}
For the radiative transfer, the Milky Way $R = 3.1$ dust model from \citet{Weingartner01} is used, with the \citet{Draine07} update. \textsc{Sunrise} allows two different treatments of the sub-resolution dust structure. In the ``default ISM'' (DISM) approach, the dust associated with cold clouds in the \citet{Springel03} sub-resolution models is ignored, whereas in the ``alternate ISM'' (AISM) approach, the total dust mass is used. The net effect of this in the radiative transfer is that in the case  of the AISM models, photons are propagated through more dust, and the actual amount of dust varies from cell to cell, because the fraction of ISM contained in cold gas depends on the local conditions. The variations in the SED shape due to a particular choice of dust sub-structure are thoroughly discussed in L14. It is important to note that the choice of the sub-structure dust model is perhaps the most significant uncertainty in the radiative transfer calculations. In this paper we use the DISM subresolution model only. The reason for this is that we are mostly interested in merger stages where star formation ins enhanced, and emission is dominated by the emission from \hii\ regions and PDRs, with little contribution from cold dust.

The emerging SEDs of these systems are computed for $\sim 100$ snapshots spanning a time range of $\sim 6~\mbox{Gyr}$. The time between snapshots is 100 Myr, although a finer time resolution (10 or 20 Myr) is used near the peaks of star formation, for example during the coalescence phase of the interactions. SEDs are obtained for seven different viewing angles. In the case of interacting systems, the galaxy pairs were put in a specific parabolic orbit with initial separations increasing with the mass of the larger galaxy. We obtain mock photometry of the simulated systems by convolving (assuming $z=0$) the resulting SEDs with the filter response of the following bands: \emph{GALEX}-FUV, \emph{GALEX}-NUV, U, B ,V (Johnson), J, H, K$_{\rm{S}}$ (2MASS), \emph{Spitzer}-IRAC 3.6~\mum, \emph{Spitzer}-IRAC 4.5~\mum, \emph{Spitzer}-IRAC 5.8~\mum, \emph{Spitzer}-IRAC~8\mum, \emph{IRAS} 12~\mum, \emph{Spitzer}-MIPS 24~\mum,  \emph{IRAS} 25~\mum, \emph{IRAS} 60~\mum, \emph{Herschel}-PACS 70~\mum, \emph{IRAS} 100~\mum, \emph{Herschel}-PACS 100~\mum, \emph{Spitzer}-MIPS 160~\mum, \emph{Herschel}-PACS 160~\mum, \emph{Herschel}-SPIRE 250~\mum, \emph{Herschel}-SPIRE 350~\mum, and \emph{Herschel}-SPIRE 500~\mum.

\section{\textsc{Chiburst}: a Bayesian Monte-Carlo fitting algorithm for star-forming galaxies}
\label{sec:fitting}

\subsection{Overview of the SED models}
We describe the statistical details of our fitting algorithm on Appendix \ref{app:fitting}. Here we briefly describe the astrophysical aspects of the models on which the fitter is based. \textsc{Chiburst} is based on the original star-forming galaxy SED models described in \citet{Dopita05, Dopita06a, Dopita06b} and \citet{Groves08}. These models compute the SED of a star-forming galaxy as the combination of three main components (each of them accounting for a normalized continuous SFR over different timescales) that can be scaled to adjust the star formation history (SFH) of each galaxy: \textit{(i)} a starburst (ionizing stars + \hii\ regions) population of young stars with a continuous SFR averaged over a period of 10Myr (SFR$_{10}$); \textit{(ii)} a population of stars formed at a constant rate between 10~Myr and 100~Myr ago (SFR$_{100}$); and \textit{(iii)} a component of very recent ($< 1$~Myr) star formation represented by Ultra-Compact \hii\ regions (UCHIIRs) responsible for dust heating  at temperatures of $\sim$300~K (SFR$_{1}$). In order to model more realistic galaxies, we have used \textsc{Starburst99} \citep{Leitherer99} to compute a fourth component \cite[additional to the prescription in][]{Groves08} and included it in our models to account for the population of even older (up to 5~Gyr) field stars, which we parametrize according to their total mass ($M_*$). This addition extends the range of galaxy types that we can study with the models to include systems where older populations significantly contribute to optical and near-infrared (near-IR) wavelengths.

The starburst component comes in two flavors that can be added linearly: a \textit{naked} \hii\ region (stars $+$ atomic gas) and an \hii\ region fully covered by a photon-dominated region (PDR) layer, which is the interface between the ionized gas in the HII region and the molecular gas birth cloud from which the cluster was formed. We interpret their linear combination as a covering fraction of PDR surrounding the \hii\ region ($f_{\rm{PDR}}$). As thoroughly described in \citet{Groves08}, the exact shape of the SED for these two components is controlled by two main parameters: the ISM metallicity and a dimensionless quantity related to the dust heating flux: the compactness parameter ($\mathcal{C}$). 

\subsection{The compactness parameter}
\label{sec:compactness}
 
The motivation to introduce the concept of compactness arises from the fact that, when dealing with highly complex systems such as starbursts, where different factors play a role in the heating of the dust, it seems rather inaccurate to describe the ISM conditions in terms of a single $T_{\rm{dust}}$ (or even in terms of a linear combination of several values of $T_{\rm{dust}}$). Instead, we can think in terms of a distribution of $T_{\rm{dust}}$ that changes in time as the starburst progresses. Rather than characterizing the ISM dust with a single $T_{\rm{dust}}$, we can do this in terms of a unique evolution of this distribution of $T_{\rm{dust}}$ as a function of time, $<T_{\rm{dust}}(t)>$. Since the latter is determined at any time by the intensity of the stellar radiation field produced by a cluster with luminosity $L_*$, at radius $R$ in a spherical nebula $<T_{\rm{dust}}(t)>$ is a function of the time-dependent dust heating flux $<L_*(t)>/<4*\pi*R(t)^2>$. Hence, models that preserve the run of $T_{\rm{dust}}$ with time should also preserve the quantity $L_∗/R^2$ averaged over time:

\begin{equation}
\label{eq:logC_original}
\log \mathcal{C} \propto \frac{<L_*(t)>}{<R(t)^2>}
\end{equation}

The compactness parameter is proportional to this ratio, and hence sets the evolution of the $T_{\rm{dust}}$ distribution with time. By assuming that $<L*(t)>$ scales with cluster mass, and by considering the time evolution of the radius and pressure of a mass-loss bubble as in the \citet{Castor75} approximation, the parameter $\mathcal{C}$ can be written in terms of the average star cluster mass $M_{\rm{cl}}$, and the average pressure of the ISM normalized by the Boltzmann' constant($P_0/k$) in a given system:

\begin{equation}
\label{eq:logC}
\log \mathcal{C}=\frac{3}{5}\log\left(\frac{M_{\rm{cl}}}{M_{\astrosun}}\right)+\frac{2}{5}\log\left(\frac{P_0/k}{\mbox{cm}^{-3}\: \mbox{K}}\right)
\end{equation}

In \S~\ref{sec:comp_vs_T} we will discuss the usefulness of the compactness parameter and how measuring this parameter is different from measuring dust temperatures from the SED.

\begin{figure*}[!th]
  \centering
  
\includegraphics[scale=0.4,angle=0]{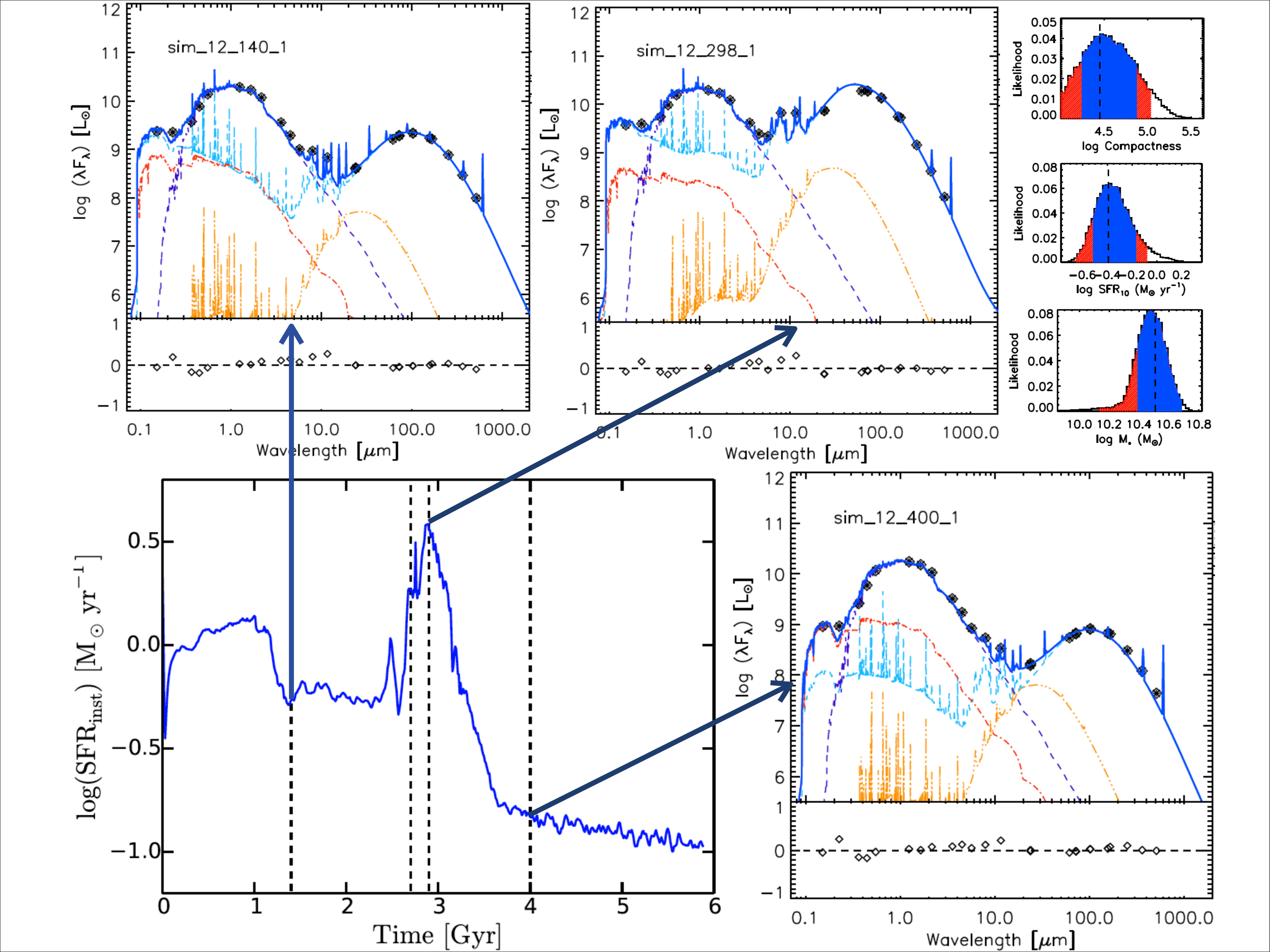} 

  \caption{The evolution of the SED during the M2-M3 interaction. The instantaneous SFR of the simulation as a function of time is plotted on the lower left corner. The dashed lines indicate particular moments along the interaction. Star formation is enhanced during the first passage at around 1~Gyr and reaches an absolute maximum at coalescence, after about 3~Gyr from the start of the simulation. Also shown are the SEDs for three of the snapshots indicated. Different SED component are color-coded as follows: \hii +PDR region (dashed cyan); 10-100 Myr population (single dot-dashed red); UCHIIRs (double dot-dashed yellow); and 5~Gyr population (dashed purple). The panels to the right are examples of the derived probability density functions (PDFs) for the model parameters for the 4~Gyr snapshot. Shaded areas corresponding to the 1$\sigma$ confidence region (blue) and the 90\% confidence region (red). The dotted line in each sub-panel corresponds to the best fit values. The labels on the top-left corner of each SED plot refer to the simulation number, snapshot number, and viewing angle.}

\label{fig:sfr_history_all}
\end{figure*}

\section{Results}
\label{sec:results}

\subsection{Fitting of the mock photometry of simulated mergers}

We have fitted the multi-wavelength photometry of the simulated interactions described in \S~\ref{sec:hydro_models} using \textsc{Chiburst}. This also includes the M4-M4 binary interaction and the gas-rich SMG simulation. As noted, the simulations cover a broad range of galactic masses, interactions stages, luminosities, and morphologies; we therefore expect them to be a fair sample of the conditions found in real galaxies. In total, we have obtained fits for over a thousand mock SEDs, that correspond to different stages along the binary interactions of the galaxies in Table~\ref{tab:progenitors}. Specifically, the snapshots fitted account for the entire interaction sequence, starting with the initial approach and ending when the merger reaches a post-coalescence passively evolving stage (total times are between 2.5~Gyr and 6~Gyr, depending on the total mass of the merger). A higher time resolution is used near the coalescence phase, and in the case of M4M4, for the entire duration of the simulation. In this paper we limit our study of the mock SEDs to a single viewing angle, but further down in this section we briefly explore the effect of viewing angle on the derived parameters. 

To illustrate the evolution of the SED along the interaction sequence for a particular simulation, Fig. \ref{fig:sfr_history_all} shows an overview of SED fits at different stages. Plotted in the lower left corner is SFR$_{\rm{inst}}$ for the M2-M3 interaction as a function of time, as calculated in the hydrodynamical simulation. SFR$_{\rm{inst}}$ increases during the first passage at around 1~Gyr after the start of the simulation and then increases abruptly during the coalescence phase, at around 3~Gyr. We have indicated three particular times along the interaction sequence that serve as representative stages of the interaction: right after the first passage, at the peak of SFR$_{\rm{inst}}$, and during the phase after coalescence. 

For the selected snapshots, we show the mock SEDs and the corresponding best \textsc{Chiburst} fits in the insets surrounding the time evolution plot. The most remarkable change in the SED as the interaction evolves is in the relative contribution of far-IR emission (coming from the \hii\ regions) to the bolometric luminosity of the system, which peaks during coalescence. The contribution of optical and near-IR emission from the oldest field stars remains rather constant along the interaction, although it increases as more stars are formed during the burst. The UV emission (to which both the \hii\ regions and the stars formed during the 100~Myr prior to a given snapshot contribute) follows the far-IR emission, although it significantly decreases in the relaxation phase after coalescence. A similar behavior is observed in all remaining simulations, with the SEDs of less massive interactions and isolated galaxies evolving much less dramatically than in the strong mergers. 

Also shown in the small insets are the probability density functions (PDFs) derived for three of the model parameters: $\log ~\mathcal{C}$, SFR$_{10}$, and $M_*$. The shaded regions in these PDFs correspond to the 1$\sigma$ (blue) and 90\% (red) confidence levels and represent the most general description of the uncertainties involved in the fitting when all possible points of the parameter space are considered. In a few cases, the best-fit values are outside the 90\% confidence region, which suggests that they can differ very significantly from the median-likelihood values. Model degeneracies, as well as the marginalization of a multi-dimensional joint probability distribution, are likely responsible for this behavior. 

The 1$\sigma$ confidence regions estimated from the PDFs are small compared to the parameter allowed ranges (typically a third or a fourth of the range), and therefore we are able to infer real differences in the parameters between simulations. One example is the SFR averaged over the last ten million years ($\mbox{SFR}_{10}$), estimated from the contribution of the \hii +PDR region to the bolometric SED. Near coalescence, \textsc{Chiburst} gives a 1$\sigma$ range for this parameter between $\sim$0.50 and $\sim$0.80~$\mbox{M}_{\odot}~\mbox{yr}^{-1}$ for the M1-M1, whereas for the M1-M2 interaction the obtained solutions range between 0.30 and 0.65~$\mbox{M}_{\odot}~\mbox{yr}^{-1}$. This relative difference in the instantaneous SFR agrees with the values predicted by the hydrodynamical simulation. Further down, in \S~\ref{sec:SFR}, we will investigate how the derived PDFs relate to the actual values from the simulations.

The SED fit plots also show how different model parameters control the fluxes in different wavelength bands: the UV is dominated by emission from stars younger than 100~Myr, either those associated with the unobscured, youngest and most massive systems in the \hii\ regions and parametrized as SFR$_{10}$ (dashed cyan line), or field A-type stars lasting 10 times longer and parametrized by SFR$_{100}$ (single-dot dashed red line); the optical and near-IR are dominated by the photospheres of the oldest field stars parametrized by $M_*$ (dashed purple line), whereas the mid-IR and far-IR emission comes almost entirely from the \hii\ regions and the PDRs with bright polycyclic aromatic hydrocarbon (PAH) emission (because no heating of diffuse dust from older stellar populations is included in the models), again parametrized by SFR$_{10}$ (the dashed light blue line again). In certain systems, there might be a non-negligible contribution to the mid-IR from UCHIIRs, parametrized by SFR$_1$ (triple-dot dashed yellow line)\footnote{In real systems, another relevant contribution to the mid-IR comes from thermal emission from the dust torus surrounding an AGN. In the simulations, AGN emission is accounted for as described in \citet{Springel05}, but \textsc{Chiburst} does not include the AGN component. We have therefore limited this study to snapshots where the contribution of AGN to the bolometric luminosity is less than 20\%. Only a few snapshots very close to coalescence in the simulated M2-M2 and M3-M3 interactions have larger contributions from the AGN. As shown in \citet{Ciesla15}, SFR and stellar mass estimates are only marginally affected by the presence of an AGN for small fractional AGN luminosities.}.

\begin{figure*}[!ht]
  \centering
  \subfigure[M0-M1]{

\includegraphics[scale=0.35,angle=90,trim=0cm 19.2cm 0cm 0cm,clip=true]{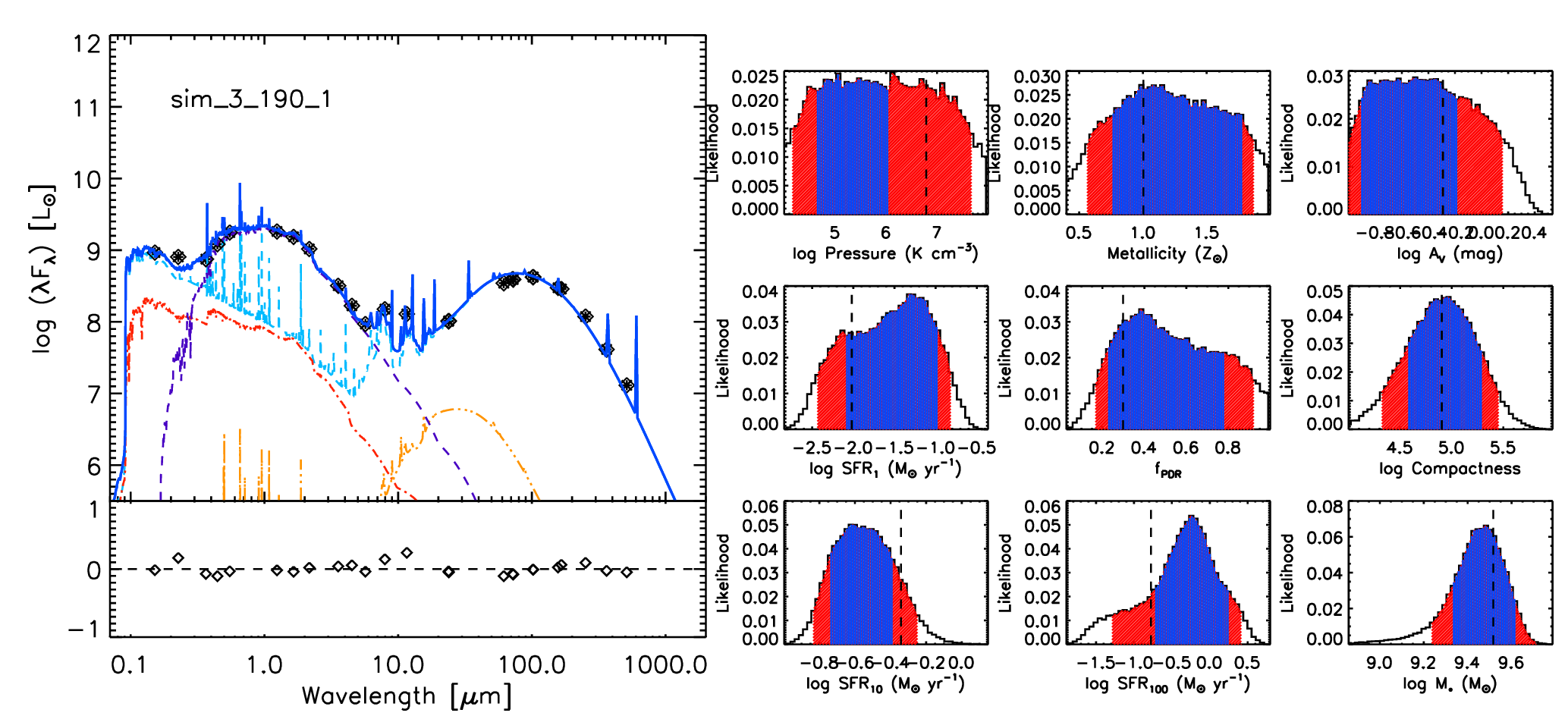}
\label{fig:fit_mass1}
}
  \subfigure[M1-M1]{
\includegraphics[scale=0.35,angle=90,trim=0cm 19.2cm 0cm 0cm,clip=true]{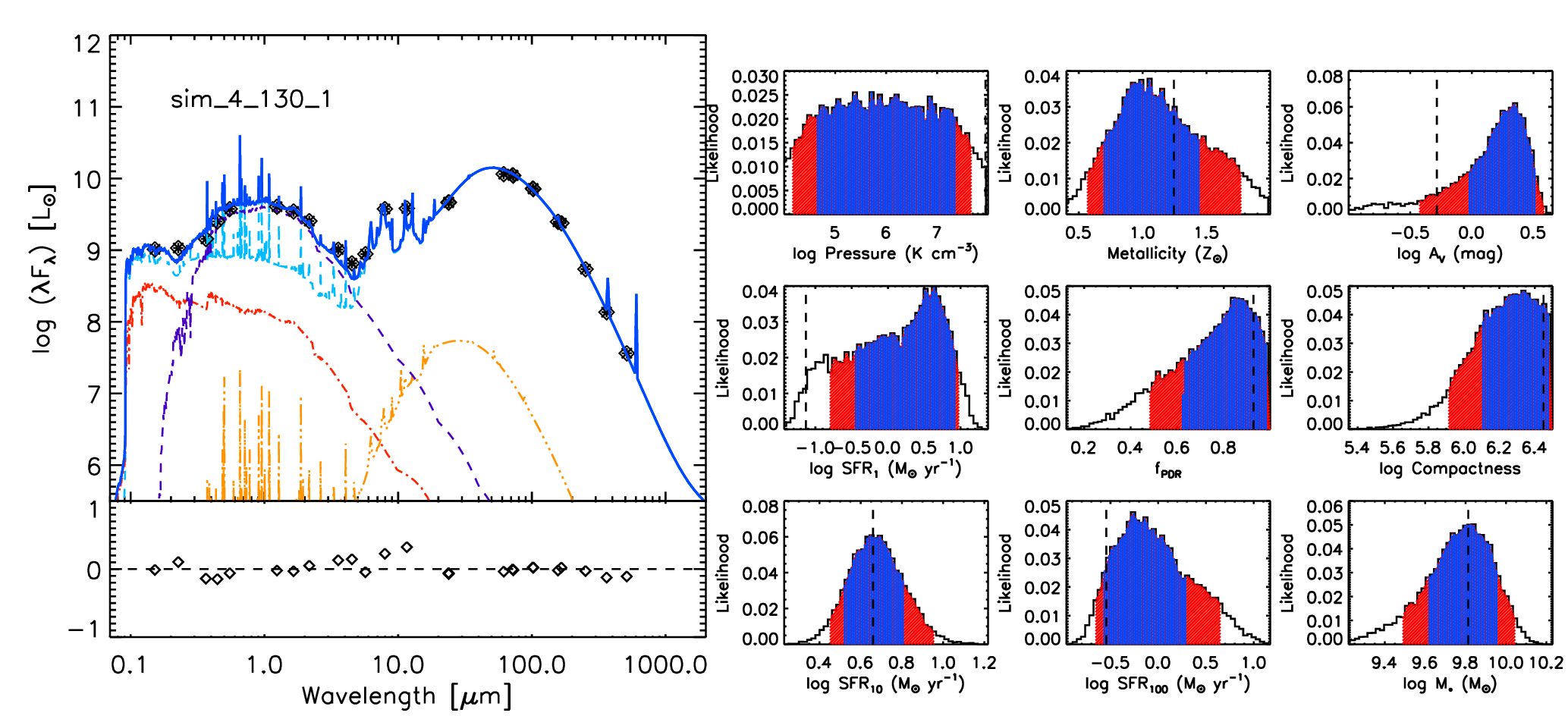}
\label{fig:fit_mass2}
}
  \subfigure[M2-M3]{
\includegraphics[scale=0.35,angle=90,trim=0cm 19.2cm 0cm 0cm,clip=true]{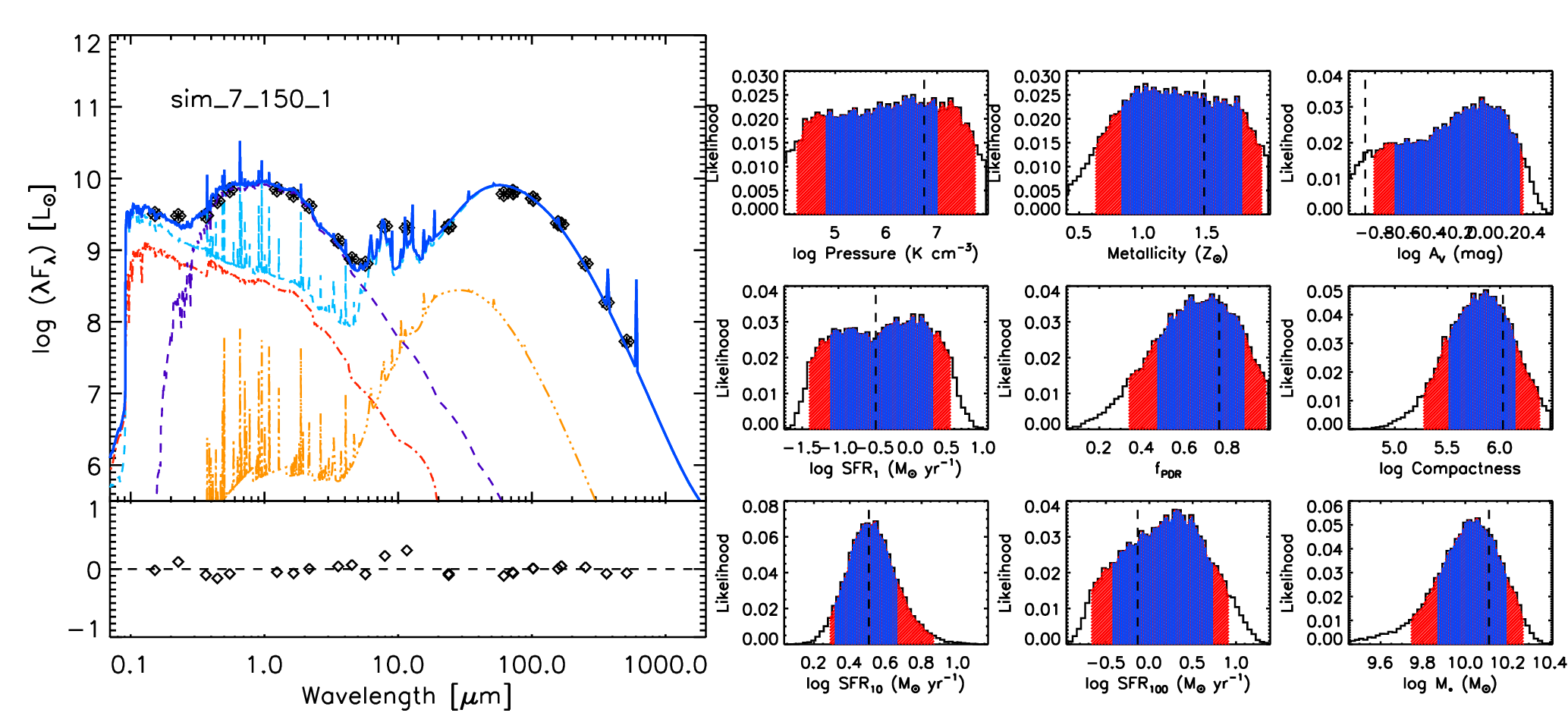}
\label{fig:fit_mass4}
}
  \caption{Fitted SEDs of three galactic interactions during coalescence, corresponding to the following pairs: \textit{(a)} M0-M1,  \textit{(b)} M1-M1, and \textit{(c)} M2-M3. The photometric data are shown as black diamonds and the best fit with \textsc{Chiburst} is shown as the blue line. Lines and color code is the same as in Fig.~\ref{fig:sfr_history_all}.}

\label{fig:fit_mass}
\end{figure*}

To illustrate how the method works for mergers of different masses, in Figs.~\ref{fig:fit_mass1}-\ref{fig:fit_mass4} we show fits to the mock SEDs of the simulations corresponding to the interaction pairs M0-M1, M1-M1, and M2-M3. In all three cases, the SEDs correspond to the coalescence phase. Typically, more massive interactions have larger bolometric luminosities, and the relative contribution of far-IR (8-1000~\mum) emission to the total luminosity tends to increase with the total mass of the interacting galaxies, an early indication that more massive interactions have larger sSFRs near the coalescence phase. 

The latter happens at first order only, because the mass ratio of the interacting pair affects the relative contribution of dust emission to the SED. The M1-M1 interaction, for example, has a larger far-IR contribution at coalescence than the M1-M2 system, which is more massive. In fact, our results show that the largest fractional far-IR luminosities happen preferentially in interacting pairs where the mass ratio is close to 1. This agrees with results from the numerical simulations of \citet{Cox08}, who show that the strength of the starburst induced by a merger decreases as the mass ratio between the progenitors increases. There is also a trend for the far-IR bump peaking at shorter wavelengths in those systems where the far-IR emission is larger \citep[see also][]{Hayward12}. We will later interpret this peak shift in terms of the compactness parameter (Eq.~\ref{eq:logC}), and will argue that this parameter and its correlation with the sSFR reveal insightful properties of the ISM physics in interacting systems.

As discussed in L14 and in \citet{Hayward15}, the viewing angle mostly affects the SED in the UV bands, where the obscuring effect of dust is larger. At these wavelengths, the typical variations in $\lambda F_{\lambda}$ are of the order of 0.25-0.5 dex, and reach a maximum immediately after coalescence, when a large amount of UV photons from young massive stars are absorbed by thick layers of dust. To study the effect of viewing angle on our derived parameters, we have fitted a post-coalescence snapshot SED from the M3-M3 simulation viewed from two different, orthogonal angles. The derived parameters PDFs do not change significantly, except perhaps for a difference of 0.1 dex in the derived value of $\log A_V$. More relevant for the present study is the fact that differences in viewing angle do not significantly affect the estimation of the compactness parameter or the SFR, because these parameters are better constrained using the mid-IR and far-IR emission, where dust obscuration is negligible or nonexistent. Moreover, we know that in the post-coalescence phase, when the effect of viewing angle is more important, the infrared indicators significantly overestimate the SFR \citep{Calzetti10, Hayward14b}, and so we interpret the post-coalescence parameters with caution.

\subsection{Validating the method: true physical parameters vs. derived parameters}
\label{sec:SFR}

\begin{figure*}[!ht]
  \centering
  \subfigure[]{
\includegraphics[scale=0.33,angle=270,trim=0cm 0cm 0cm 4.5cm,clip=true]{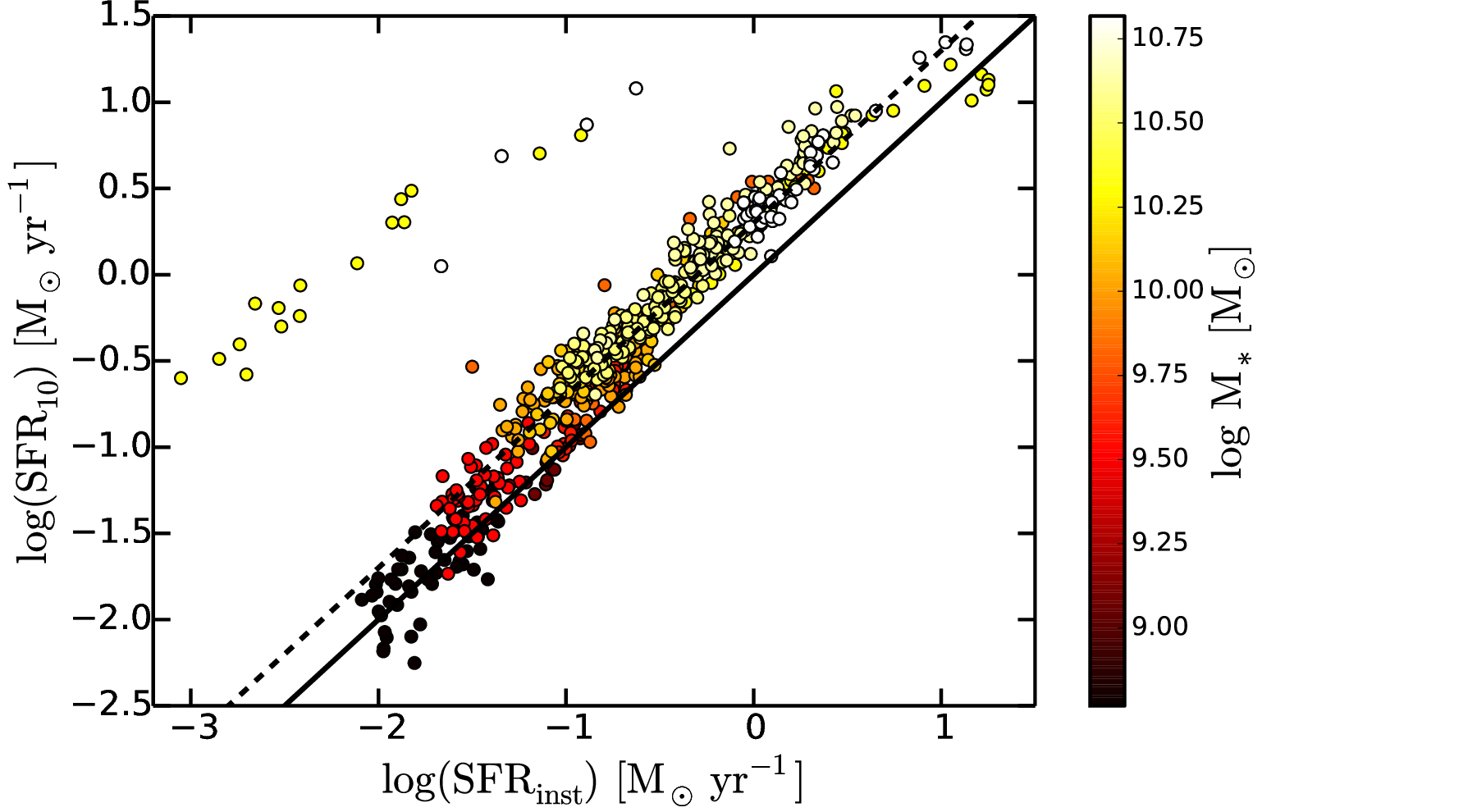}
\label{fig:SFR_comp}
}\hspace{0.2cm}
  \subfigure[]{
\includegraphics[scale=0.33,angle=270,trim=0cm 0cm 0cm 4.5cm,clip=true]{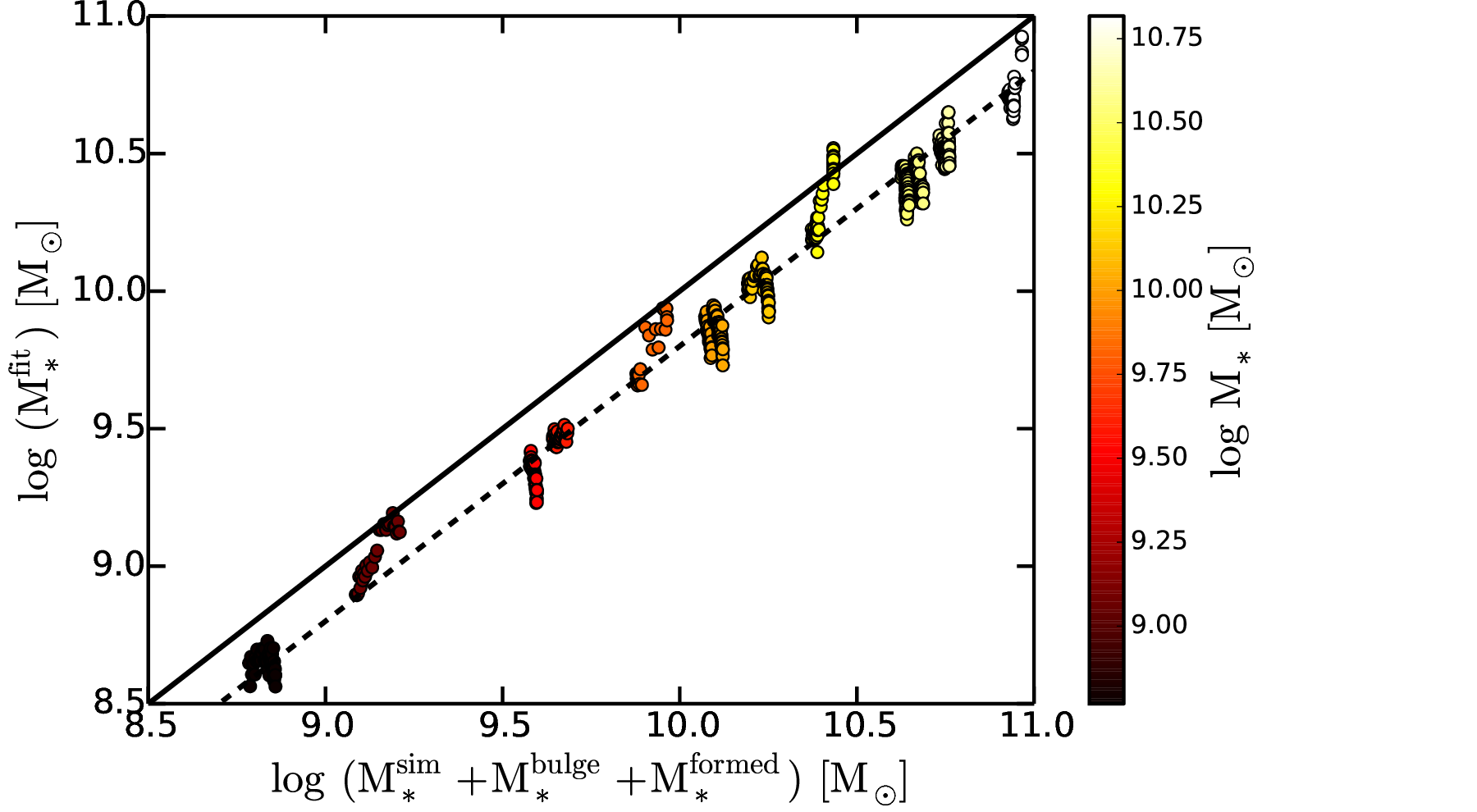}
\label{fig:Md_comp}
}

  \caption{\emph{(a)} Comparison between the instantaneous SFR$_{\rm{inst}}$ (from the hydrodynamical simulations) and the \textsc{Chiburst} derived values of SFR$_{10}$ (peak of the PDF), for the complete set of the L14 simulations (including isolated galaxies and mergers). The solid line indicates the one-to-one correlation, whereas the dashed line is shifted upwards by a factor of 2. \emph{(b)} Comparison between the total stellar mass in each simulation (including initial disk and bulge masses and the mass of stars formed from the beginning of the simulation) and the estimated parameter $M_*$ from \textsc{Chiburst} (peak of the PDF). The solid line indicates the one-to-one correlation, whereas the dashed line is shifted downwards by a factor of 1.25. In both panels, the points are color-coded by total stellar mass.}

\label{fig:true_vs_derived}
\end{figure*}

In this section we compare the parameter values derived from the fitting of mock SEDs for the original L14 simulations with the \textit{true} values from the hydrodynamical simulations. As mentioned, in the simulations unresolved star formation is accounted for by assuming that gas particles with densities above certain threshold form stars according to the volume-density-dependent S-K law. The resulting star particles are assigned \textsc{Starburst99} SEDs with a Kroupa initial mass function (IMF), which are the input for the radiative transfer code. Since \textsc{Chiburst} assumes the same IMF, we assume the \textit{true} SFR to be the instantaneous SFR from the simulations. In Fig.~\ref{fig:SFR_comp} we show the correlation between SFR$_{\rm{inst}}$ and the \textsc{Chiburst}-derived value of SFR$_{10}$.

\textsc{Chiburst} overestimates the instantaneous SFRs, but not dramatically: for the most massive interactions (M2-M3, M3-M3), we obtain SFRs that are an average factor of 2 above the true values, whereas for the less massive systems SFR$_{\rm{inst}}$ and SFR$_{\rm{10}}$ are in good agreement within the 1$\sigma$ uncertainties derived from the PDFs. The outliers of this correlation, significantly above the majority of points (up to a factor of 100), correspond to the post-coalescence phase of the most massive mergers, where star formation is abruptly quenched. One possible reason for the post-coalescence overestimation of the SFR is heating from stellar populations older than 10~Myr \citep{Hayward14b, Groves12}, which is not accounted for in the models. When dust heating from older stellar populations is included in the SED modeling, the SFR can be recovered much more accurately in the post-coalescence phase \citep{Hayward14b}. After coalescence, the fraction of recently formed stars decreases dramatically, and this effect becomes more important. Note, however that the correlation spans over a range of 3 orders of magnitude in $\mbox{SFR}_{\rm{inst}}$. The SMG simulation is also in good agreement with this correlation, which indicates that \textsc{Chiburst} can determine the true values of SFR within a factor of 2 over a dynamic range covering almost five orders of magnitude.

We also compare the stellar mass in each simulation with our estimation of the $M_*$ parameter from \textsc{Chiburst}. In Fig.~\ref{fig:Md_comp} we show this comparison for the original L14 simulations. We slightly underestimate the total stellar mass, obtained by summing up the initial disk and bulge stellar masses, and the mass of stars formed during the simulation. We obtain stellar masses that are a factor of 1.25 or less below the true stellar masses, with better agreement near the coalescence phase. The combined effect of simultaneously overestimating the SFR and underestimating the stellar mass implies that our derived values for the sSFR ($\mbox{sSFR}\equiv \mbox{SFR}_{\rm{inst}}/\mbox{M}_*$) are a factor of 2.5 (or 0.4 dex) above the true sSFR values. 

These is a systematic effect and therefore does not affect the conclusions we make below regarding \emph{relative} correlations between model parameters. In general, the uncertainty in the determination of the sSFR is method-dependent. By collecting published values of the sSFR as a function of stellar mass and redshift using various methods, \citet{Behroozi13} have estimated the uncertainty in sSFR due to the use of different techniques. They show that such uncertainty varies from 0.3 dex to 0.4 dex for stellar masses between $10^{10.5}~\rm{M}_{\odot}$ and $10^{9.5}~\rm{M}_{\odot}$. Our derived values for sSFR are therefore within the uncertainty associated with the use of a particular method.

\subsection{Fitting the SEDs of real galaxies}
Using \textsc{Chiburst}, we have fitted the multi-wavelength SEDs of the galaxies listed in Table~\ref{tab:obs_list}. For most of these systems, the wavelength coverage is similar to that of the mock photometry presented in the last section, although some of the bands are missing for specific galaxies. The intermediate-$z$ LIRGs appear heavily obscured, and therefore no \emph{GALEX} detections are available. For these obscured systems, however, the rest-frame UV emission that for local systems falls within the \emph{GALEX} bands is redshifted into the optical, making the lack of \emph{GALEX} data a less severe problem for our purposes.

\begin{figure*}[!ht]
  \centering
  \subfigure[Local merger - M82]{

\includegraphics[scale=0.35,angle=90,trim=0cm 19.2cm 0cm 0cm,clip=true]{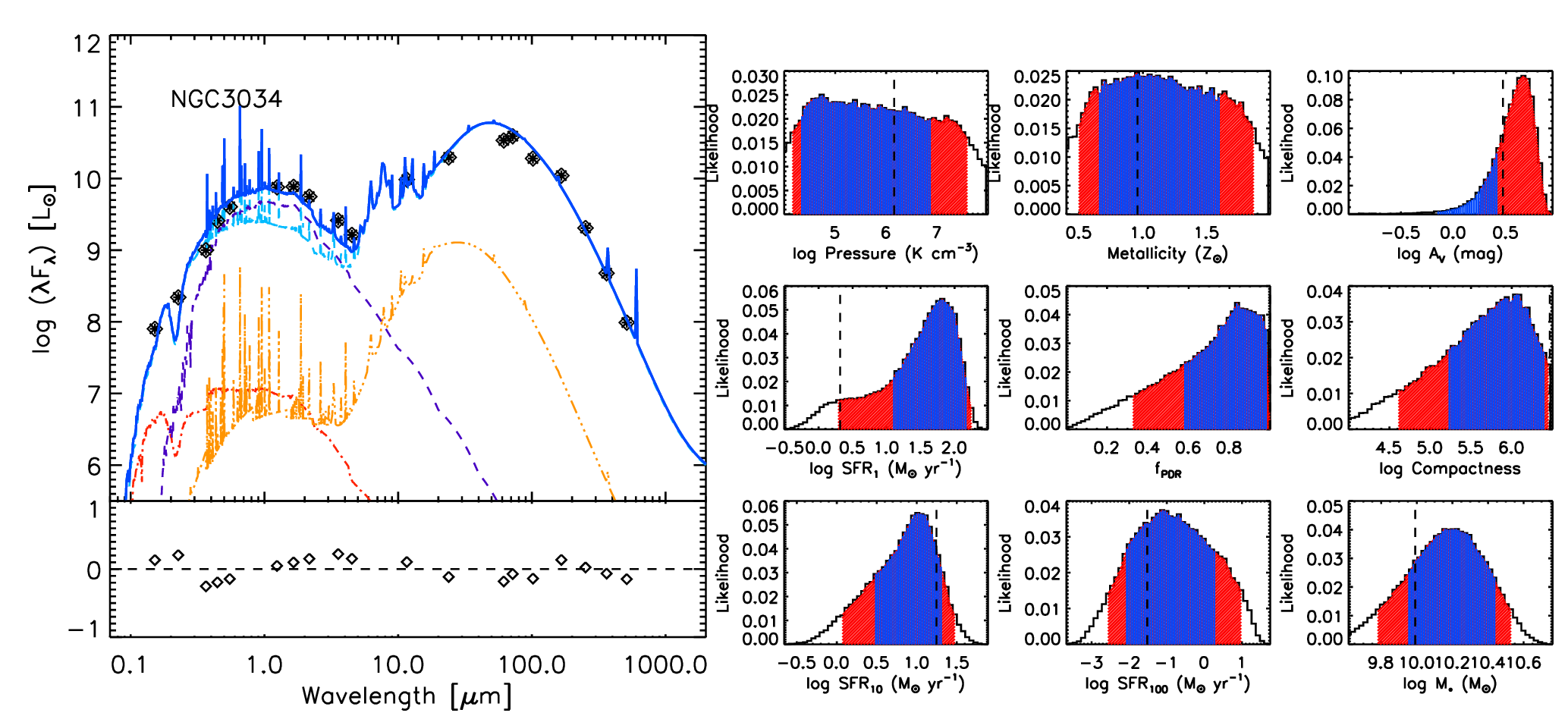}
\label{fig:fit_M82}
}
  \subfigure[Local luminous late-type merger - NGC~6090 ]{
\includegraphics[scale=0.35,angle=90,trim=0cm 19.2cm 0cm 0cm,clip=true]{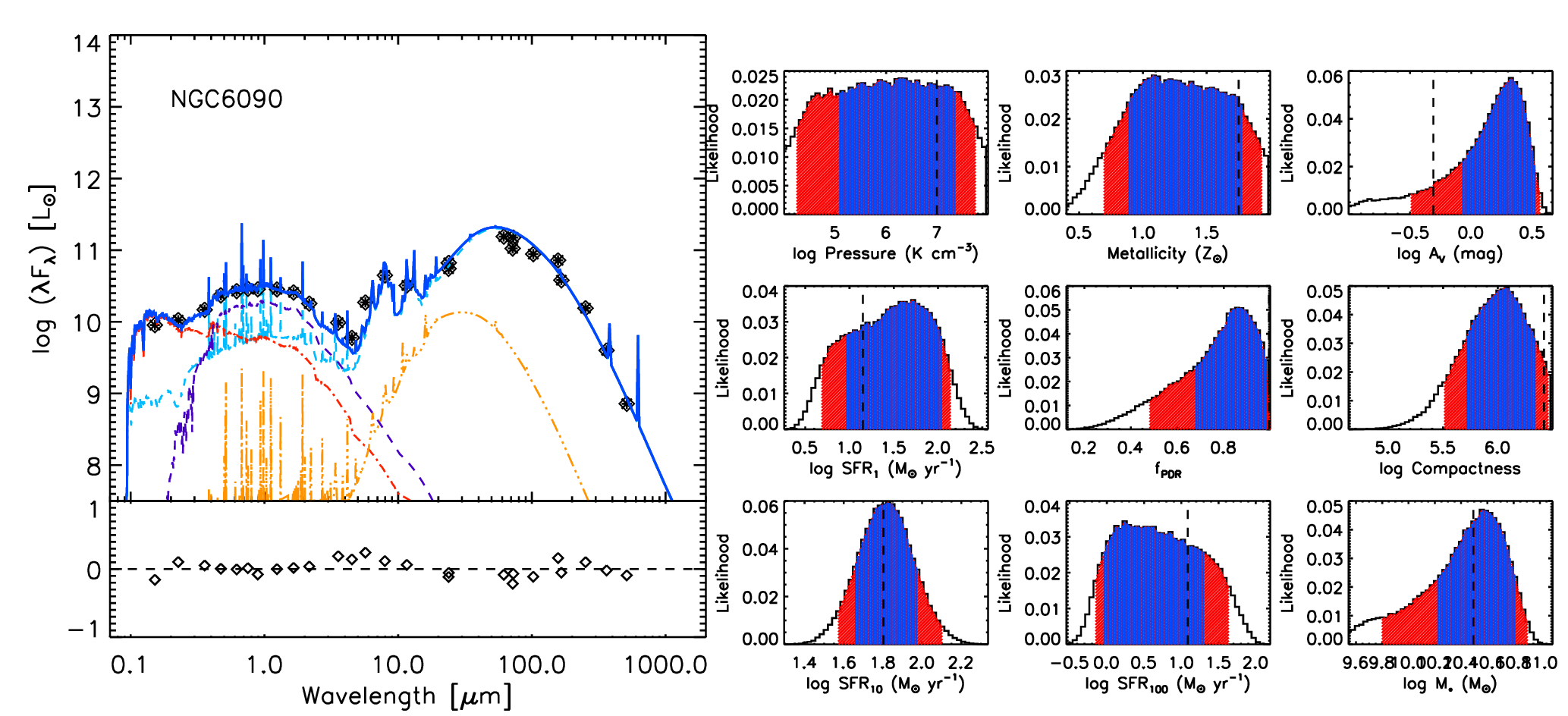}
\label{fig:fit_NGC6090}
}
  \subfigure[LIRG at z = 0.248 - CDFS2]{
\includegraphics[scale=0.35,angle=90,trim=0cm 19.2cm 0cm 0cm,clip=true]{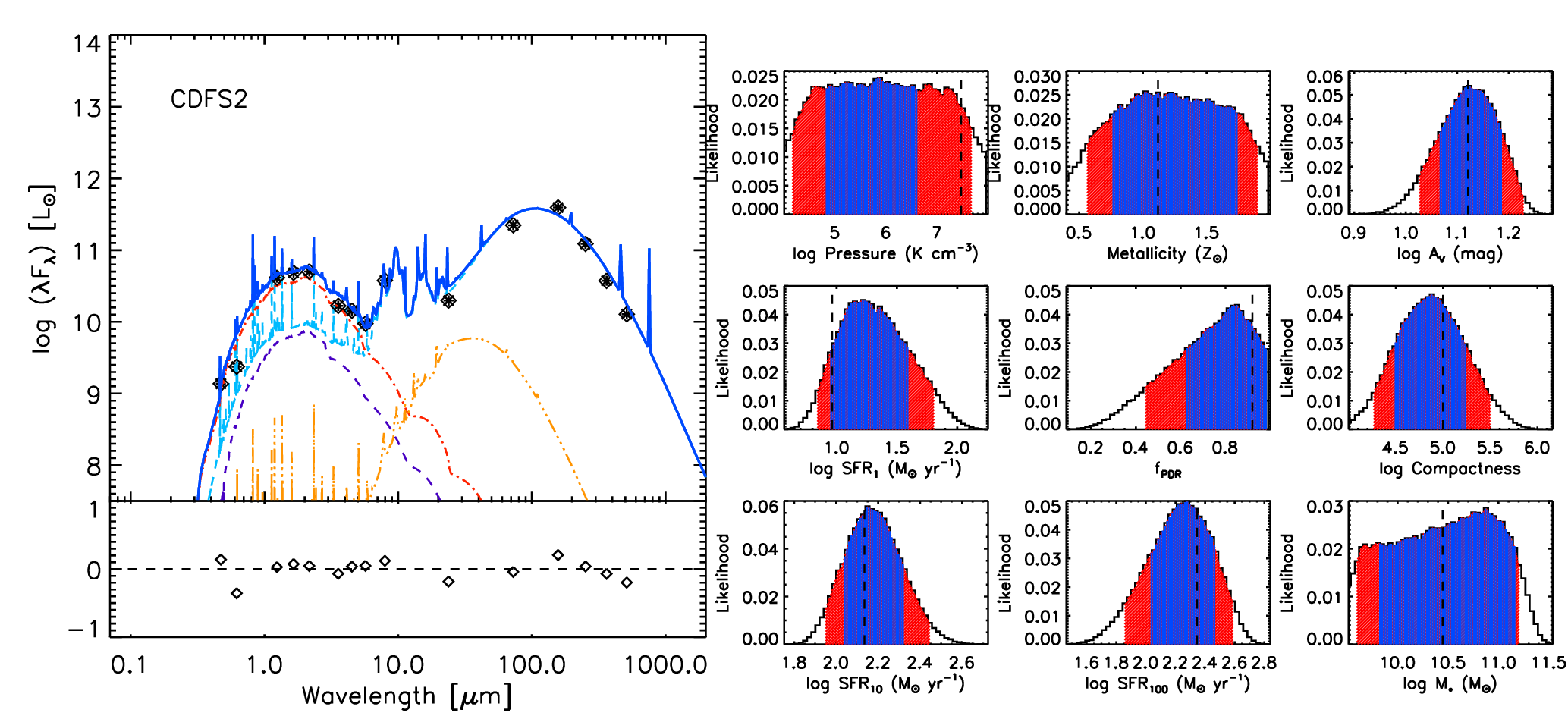}
\label{fig:fit_CDFS2}
}
  
  \caption{Example fits for each of our galaxy types. (a): A local merger; (b): A local luminous late-type merger; and (c): A LIRG at $z = 0.248$. Color code for the different components and PDFs is the same as in Fig.~\ref{fig:fit_mass}.}

\label{fig:fit_obs}
\end{figure*}

In Fig.~\ref{fig:fit_obs} we show example fits for representative galaxies in each of our three groups: local, average luminosity interactions, local LIRGs, and the intermediate-$z$ LIRGs from \citet{Magdis14}. In most cases, residuals of the fit are within $\pm 0.2$~dex over the entire wavelength range. One exception is BOOTES1 which has a flat, featureless mid-IR spectrum. Our hydrodynamical simulations, as well observational evidence from IRS observations of embedded active nuclei \citep{Imanishi07}, suggest that such flat spectrum is due to a significant contribution from mid-IR AGN thermal emission to the bolometric luminosity \citep[]{Snyder13}. None of the other systems appears to have a mid-IR spectrum compatible with a significant contribution from an AGN. As we have mentioned before, minor AGN contributions can slightly affect the SFR estimates, but not dramatically.

\begin{figure*}[!ht]
  \centering
  \subfigure[]{
\includegraphics[scale=0.32,angle=270,trim=4.5cm 1.4cm 4.5cm 1.4cm,clip=true]{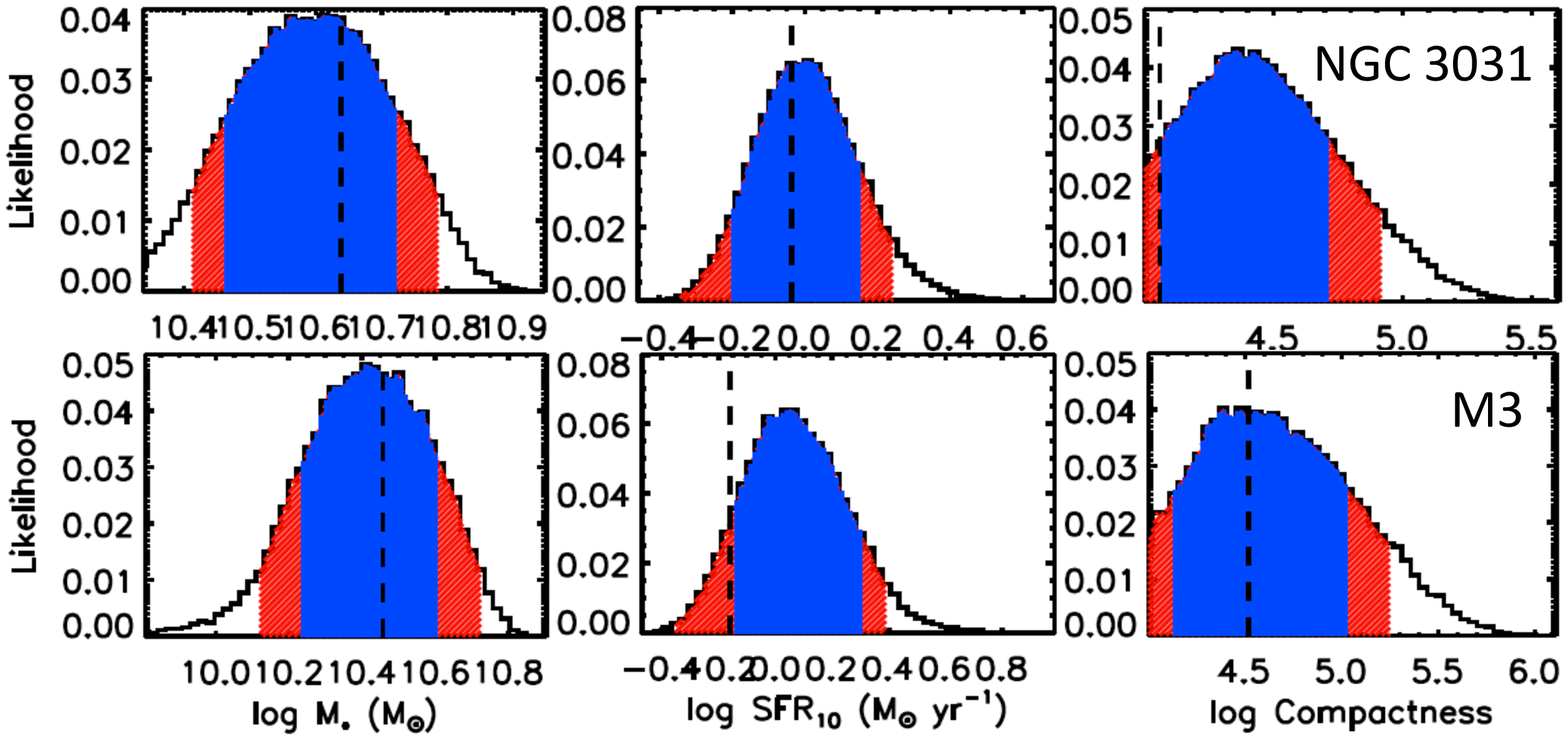}
\label{fig:comp1}
}\hspace{0.2cm}
  \subfigure[]{
\includegraphics[scale=0.32,angle=270,trim=4.5cm 1.4cm 4.5cm 1.4cm,clip=true]{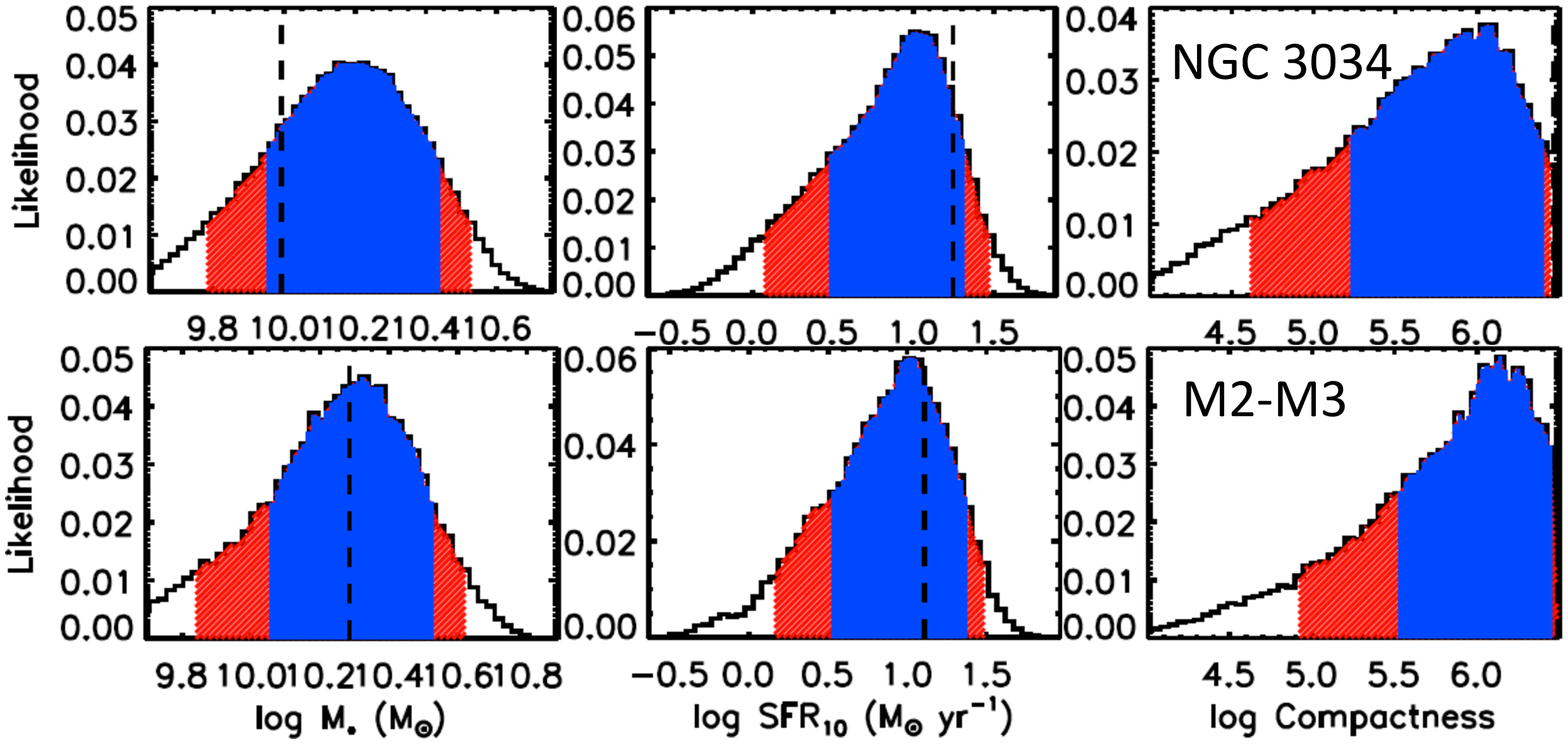}
\label{fig:comp2}
}

  \caption{Derived PDFs for observed and simulated systems at two different stages of the interaction. \emph{(a)} An almost undisturbed disk galaxy (NGC~3031) has very similar parameters as our simulated isolated disk M3. \emph{(b)} A late-stage merger (NGC~3034) reveals similar derived parameters as those for the M2-M3 interaction near coalescence. In particular, the two types of systems show here are in opposite ends of the compactness range.}

\label{fig:comp_PDFs}
\end{figure*}

Table~\ref{tab:results} lists the \textsc{Chiburst}-derived parameters for the observed systems. The range of stellar masses and SFRs that we obtain for the local interactions fall within the ranges spanned by these same parameters in the simulated interactions (see \S~\ref{sec:MS} below), which allows a comparison between the observed systems and corresponding simulated stages of the interactions. Not surprisingly, observed systems that are in later stages of the interaction (i.e., strongly interacting, morphologically disturbed systems such as NGC~3034 and NGC~3690) have SFRs and stellar masses similar to those of the most massive simulated interactions near the coalescence phase, whereas observed systems in early phases of the interactions usually have lower SFRs for similar masses, and have similar derived parameters as the early stages of the simulated interactions. Such is the case for M51a and NGC~3031. To illustrate this association between observed and simulated systems, in Fig.~\ref{fig:comp_PDFs} we show a comparison of the derived probability distributions for observed and simulated systems.

\begin{deluxetable*}{ccccccccc}
\tablecolumns{9}
\tablecaption{Fitting results}

\tablehead{
  \colhead{Galaxy} &
  \colhead{Type} &
  \colhead{$\log$~SFR$_{10}$} &
  \colhead{$\log$~SFR$_{100}$} &
  \colhead{$\log$~$M_{*}$} &
  \colhead{$\log$~SFR$_{1}$} &
  \colhead{$F_{\rm{PDR}}$} &
  \colhead{$\log\, \mathcal{C}$} &
  \colhead{$\log\, A_V$} \\
  \colhead{} & 
  \colhead{} & 
  \colhead{$[\mbox{M}_{\odot}\, \mbox{yr}^{-1}]$} & 
  \colhead{$[\mbox{M}_{\odot}\, \mbox{yr}^{-1}]$} & 
  \colhead{$[\mbox{M}_{\odot}]$} & 
  \colhead{$[\mbox{M}_{\odot}\, \mbox{yr}^{-1}]$} & 
  \colhead{} & 
  \colhead{} &
  \colhead{[mag]} 
}
\startdata

 NGC~2976 & Local merger  & -0.610  & -1.143  & 9.188  & -1.714  & 0.67  & 4.88  & -0.357 \\
 NGC~3031 & Local merger  & 0.008  & 0.113  & 10.611  & -0.936  & 0.75  & 4.36  & -0.004 \\
 NGC~3034 & Local merger  & 1.007  & -1.124  & 10.154  & 1.819  & 0.84  & 6.04  & 0.665 \\
 NGC~3185 & Local merger  & -0.120  & -0.761  & 10.073  & -0.733  & 0.81  & 5.02  & -0.057 \\
 NGC~3187 & Local merger  & -0.005  & -0.160  & 9.593  & -0.921  & 0.71  & 4.72  & -0.183 \\
 NGC~3190 & Local merger  & 0.263  & -1.400  & 10.652  & -0.617  & 0.92  & 4.82  & 0.367 \\
 NGC~3395/3396 & Local merger  & 0.848  & 0.962  & 10.144  & 0.154  & 0.62  & 5.55  & -0.463 \\
 NGC~3424 & Local merger  & 0.650  & -0.835  & 10.289  & -0.000  & 0.96  & 5.54  & 0.275 \\
 NGC~3430 & Local merger  & 0.615  & 0.355  & 10.285  & -0.749  & 0.70  & 5.27 & -0.123 \\
 NGC~3448 & Local merger  & 0.458  & 0.232  & 9.935  & -1.035  & 0.79  & 5.32  & -0.028 \\
 UGC~6016 & Local merger  & -1.414  & -0.438  & 8.140  & -2.375  & 0.82  & 4.79  & -0.722 \\ 
 NGC~3690/IC694 & Local merger  & 2.213  & 0.622  & 10.718  & 1.953  & 0.95  & 6.36  & 0.374 \\
 NGC~3786 & Local merger  & 0.382  & -0.456  & 10.331  & 0.274  & 0.81  & 5.17  & -0.032 \\
 NGC~3788 & Local merger  & 0.271  & -0.254  & 10.239  & -0.401  & 0.81  & 4.72  & -0.263 \\
 NGC~4038/4039 & Local merger  & 1.405  & 0.939  & 10.888  & 0.476  & 0.81  & 5.49  & -0.390 \\
 NGC~4618 & Local merger  & -0.556  & 0.150  & 9.300  & -1.487  & 0.83  & 4.90  & -0.768 \\
 NGC~4625 & Local merger  & -0.985  & -0.798  & 8.891  & -2.538  & 0.78  & 4.98  & -0.305 \\
 NGC~4647 & Local merger  & 0.320  & 0.323  & 10.175  & -0.793  & 0.87  & 5.28  & 0.383 \\
 M~51a  & Local merger  & 0.869  & 0.429  & 10.571  & -0.010  & 0.76  & 5.14  & -0.213   \\
 M~51b  & Local merger  & -0.081  & -1.031 & 10.213  & -0.619  &  0.85  & 5.10  & 0.263   \\
 NGC~5394 & Local merger  & 1.140  & 0.346  & 10.428  & 1.206  & 0.90  & 5.84  & 0.472 \\
 NGC~5395 & Local merger  & 1.204  & 0.513  & 11.089  & 0.298  & 0.84  & 4.94  & -0.058 \\
 M~101  & Local merger  & 0.741  & 0.847  & 10.428  & -0.107  & 0.53  & 4.94  & -0.853  \\
 NGC~5474 & Local merger  & -1.129  & -0.300  & 8.976  & -1.996  & 0.81  & 4.49  & -0.843 \\
 \hline
 NGC~2623  & Local LIRG  & 1.681  & 0.385 & 10.352  & 0.990  & 0.89  & 5.70  & 0.576 \\ 
 UGC~4881  & Local LIRG  & 1.957  & 0.361 & 11.017  & 0.979  & 0.95  & 5.55  & 0.463 \\
 VV~283  & Local LIRG  & 1.813  & 0.305 & 10.702  & 0.820  & 0.96  & 5.77  & 0.533 \\
 Mrk~273  & Local LIRG  & 2.236  & 0.559 & 10.697  & 1.991  & 0.96  & 6.48  & 0.534 \\ 
 VV~705  & Local LIRG  & 2.075  & 0.516 & 10.861  & 2.297  & 0.91  & 6.13  & 0.428 \\
 NGC~6090  & Local LIRG  & 1.827  & 0.240 & 10.557  & 1.720  & 0.86  & 6.08  & 0.324 \\ 
 \hline
 ELAISS  & Interm. $z$ LIRG  & 2.085  & 2.319 & 11.081  & 1.328  & 0.80  & 4.70  & 1.016 \\ 
 CDFS2  & Interm. $z$ LIRG  & 2.149  & 2.278 & 10.876  & 1.221  & 0.85  & 4.88  & 1.123 \\ 
 CDFS1  & Interm. $z$ LIRG  & 2.355  & 2.201 & 11.160  & 1.719  & 0.82  & 4.81  & 0.931 \\  
 SWIRE4  & Interm. $z$ LIRG  & 2.250  & 1.846 & 10.777  & 1.360  & 0.84  & 5.53  & 0.756 \\
 SWIRE5  & Interm. $z$ LIRG  & 2.609  & 1.720 & 11.590  & 1.709  & 0.85  & 5.21  & 0.608 \\
 SWIRE2  & Interm. $z$ LIRG  & 2.362  & 1.703 & 11.417  & 1.633  & 0.82  & 4.36  & 1.018 \\
 SWIRE7  & Interm. $z$ LIRG  & 2.879  & 1.845 & 11.142  & 2.146  & 0.91  & 5.49  & 0.464 \\
 BOOTES2  & Interm. $z$ LIRG  & 2.237  & 1.599 & 10.893  & 1.335  & 0.84  & 5.28  & 0.626 \\ 
 BOOTES1  & Interm. $z$ LIRG  & 3.187  & 1.671 & 11.262  & 3.755  & 0.84  & 6.09  & 0.868  
\enddata

\label{tab:results}
\end{deluxetable*}

All of the local and intermediate-$z$ LIRGs are outside the range of sSFRs covered by the binary interactions for the same stellar masses (Fig.~\ref{fig:ms_all}) (i.e., they are outliers of the MS), but they appear to be somewhere in between the region of the parameter space covered by these progenitors and the region covered by the gas-rich SMG simulation. Their compactness values are also low relative to those of local interactions with similar sSFRs. This is relevant in the context of the current discussion regarding whether the enhanced sSFRs of intermediate- and high-$z$ galaxies is due to higher gas fractions in galaxies earlier in cosmic history \citep[e.g.][]{Daddi10a, Tacconi13}, or to higher efficiencies in the conversion of gas into stars, likely driven by mergers \citep[e.g.][]{Daddi09,Frayer08}. Here we will quantify the effect of both interactions and gas content in rising a particular galaxy above the MS.

\subsection{A correlation between compactness and sSFR.}
\label{sec:compact_correl}

As originally defined in \citet{Groves08} and \S~\ref{sec:compactness}, the compactness parameter ($\mathcal{C}$) is related to the internal conditions of the ISM, specifically to the ISM pressure and the radiation field strength of individual \hii\ regions. Despite the fact that the hydrodynamical simulations considered here do not resolve the small-scale structure of the ISM and simplify its physics, the self-similar structure of the ISM implies that an analogous interplay between radiation strength and pressure takes place at the scales resolved by our simulations. Also, the sub-resolution models that describe the ISM physics at smaller scales are chosen within the simulations according to the resolved properties of the ISM for a given particle. We therefore expect $\mathcal{C}$ to be a meaningful description of the large scale geometry of the dust with respect to the stars, and hence a reasonable probe of the global galaxy compactness. 

\begin{figure*}[ht]
  \centering
  \subfigure[]{
\includegraphics[scale=0.4,angle=0,trim=0.3cm 0cm 0.1cm 0cm,clip=true]{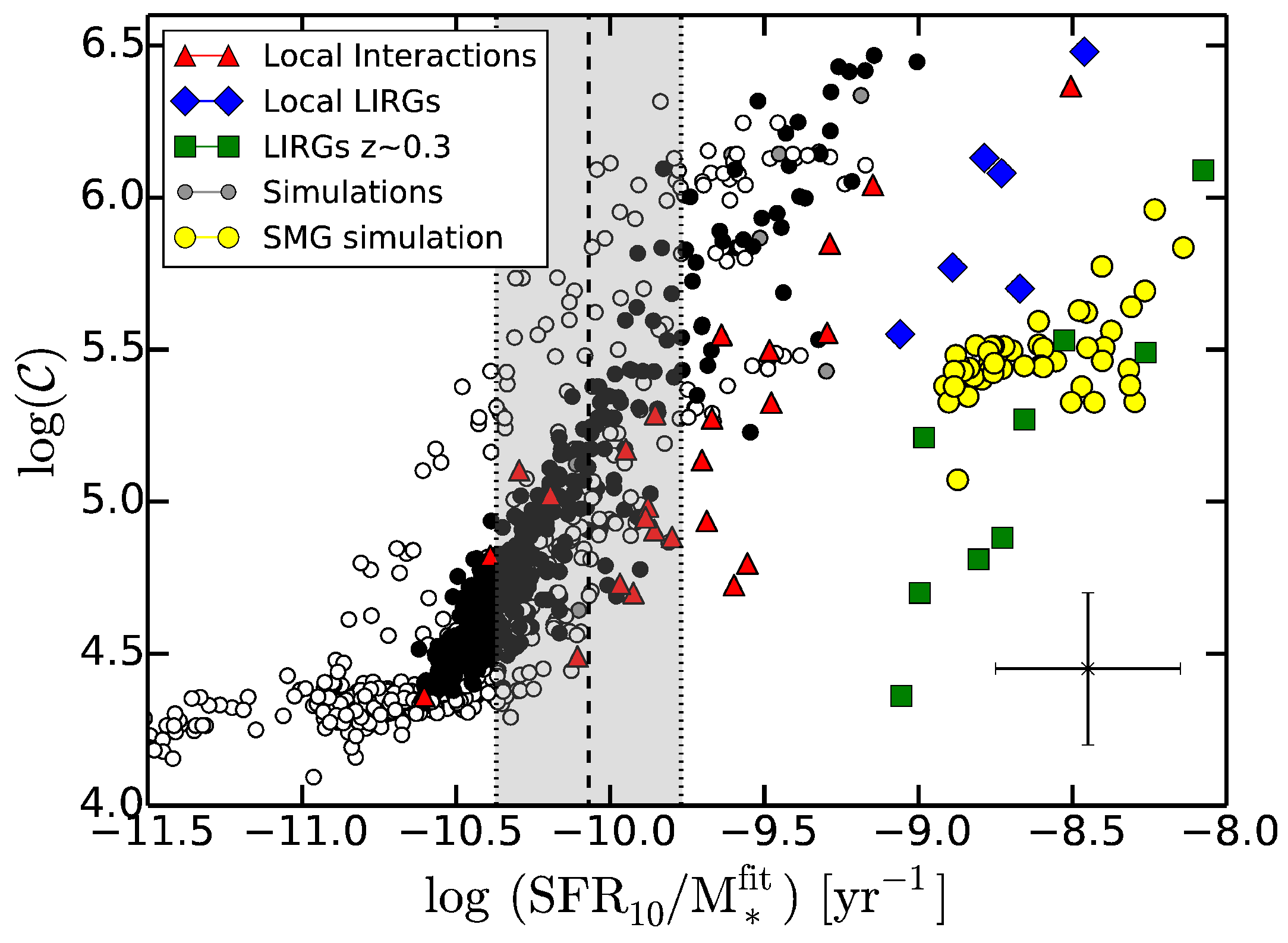}
\label{fig:sSFR_logC_a}
  }
  \subfigure[]{
\includegraphics[scale=0.4,angle=0,trim=0.3cm 0cm 0.1cm 0cm,clip=true]{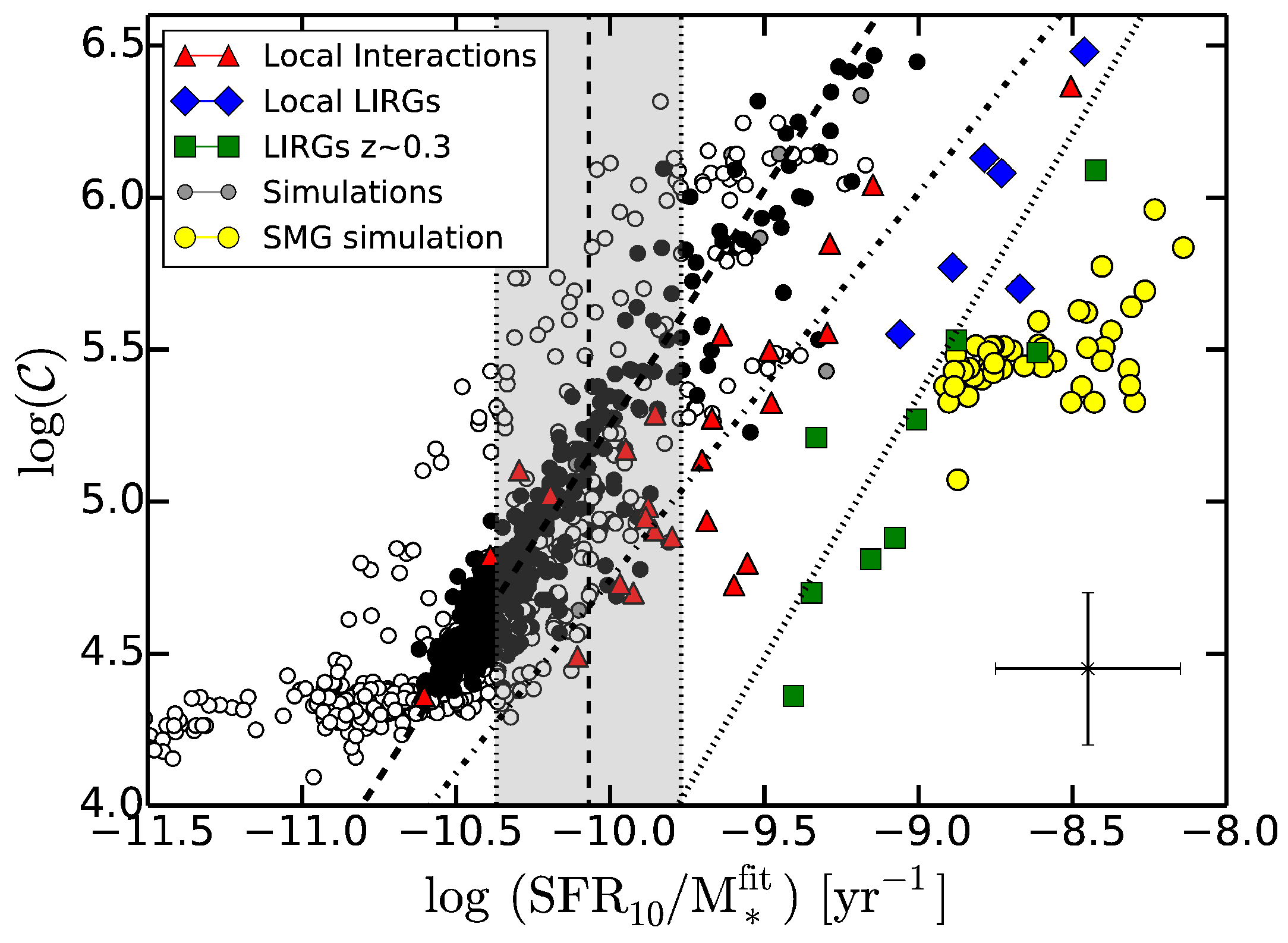}
\label{fig:sSFR_logC_b}
  }

  \caption{\emph{(a)} The logarithm of the compactness parameter ($\log \mathcal{C}$) plotted against the logarithm of the derived sSFR for all simulated and observed systems. Simulations are shown as filled (before coalescence) and empty (after coalescence) circles, each circle corresponding to a different snapshot along the interaction sequence. Also shown are the local interactions (red triangles), local LIRGs (blue diamonds), and intermediate-$z$ LIRGs (green squares). The large yellow circles correspond to the SMG simulation. The location of the $z=0$ MS according to the \citet{Elbaz11} parametrization is indicated by the vertical dashed line, and the shaded area corresponds to a 0.3~dex scatter of the MS. Typical errors in the estimated parameters are indicated in the bottom-right corner. \emph{(b)} The same plot, but with the intermediate-$z$ galaxies shifted to match the sSFR that they would have in the Local Universe, according to the same parametrization. Diagonal lines correspond to linear fits to the pre-coalescence simulations (dashed line), the local interactions (dot-dashed line) and the local and intermediate-$z$ LIRGs plus NGC~3690, which is the only LIRG in the L13 sample (dotted line). The parameters of each fit are described in the text.}

\label{fig:sSFR_logC}
\end{figure*}

Here we characterize observed and simulated systems according to their $\mathcal{C}$ parameter in order to study whether the geometry and heating conditions of dust in galaxies change as a function of particular stages along the interaction sequence. Fig.~\ref{fig:sSFR_logC_a} shows how the derived $\mathcal{C}$ values for observed and simulated galaxies correlate with the derived sSFRs. The plotted values correspond to the maximum of the marginalized PDF for each parameter. Simulations of the isolated and interacting progenitor disk galaxies (M0 to M4) at different snapshots are marked with small circles, both before (filled circles) and after (empty circles) coalescence, whereas the SMG simulation is indicated by the large yellow circles. Following the classification of Table~\ref{tab:obs_list}, we have separated the observed systems into local interactions (red triangles), local LIRGs (blue diamonds), and intermediate-$z$ LIRGs (green squares).

Prior to coalescence\footnote{Using the same suite of simulations as the one used here, \citet{Hayward14b} show that mid-IR indicators can significantly overestimate the SFR after coalescence. As mentioned above, our models do not include heating of the dust from stars older than 10~Myr, and it is therefore not surprising that the empty circles in Fig.~\ref{fig:sSFR_logC_a} depart from the correlation found.}, all simulated systems, regardless of total mass or gas fraction, appear to fall along a single power law of the form $\log \mathcal{C} \propto \log (\rm{sSFR})^{\alpha}$. This correlation between compactness and sSFR provides a physical framework that can be used to explain the correlation found by \citet{Magnelli14} for $z < 2$ galaxies between sSFR and $T_{\rm{dust}}$. In fact, \citet{Magnelli14} argue that at a given redshift, the $T_{\rm{dust}}$-sSFR correlation implies that galaxies situated above the MS are dominated by star-forming regions exposed to high radiation fields. Elaborating along similar lines, \citet{Magdis12} argue that the mean radiation field $<U>$ is the only parameter controlling the shape of the SED of high redshift galaxies. The parametrization presented here implies that, in fact, at scales larger than a few hundred parsecs, dust has a more compact geometry in outliers relative to MS galaxies. The latter can be an effect of a higher radiation field due to more massive star clusters, an increased pressure field due to the merger, or both. 

In Fig.~\ref{fig:sSFR_logC_a}, observed local interactions follow a trend that is similar in slope and normalization (see Table \ref{tab:linear_fits}) to that of the simulations, an indication that the latter constitute a reasonable control group on which we can test our assumptions on the star-forming properties of galaxies. This agreement between models and observations is even more remarkable when the morphology of the observed systems is considered. Isolated disks and early interactions such as NGC~3031 and NGC~5474 typically have low $\log~\mathcal{C}$ values, but compactness gradually increases as we go towards late stage mergers, with the highest $\log~\mathcal{C}$ values derived for strongly disturbed systems such as NGC~3034. That local interactions do not exactly overlap with the simulations in this diagram can be attributed to the fact that the latter, although designed to reproduce a broad range of galaxy types, are nevertheless imperfect constructions, simplify the ISM physics, and span a relatively small parameter space in terms of total gas masses and gas fractions. 

More interesting is the fact that both local and intermediate-$z$ LIRGs follow the same correlation, although the normalization constant in each case is different with respect to the local systems. Furthermore, both simulations and observations are consistent with late stage interactions having elevated values of compactness, which implies that the interaction stage of an unresolved galaxy can be inferred from the properties of its SED. This is consistent with L14, where the authors claim that the SED shape can be used to discriminate between coalescence and non-coalescence systems. The tight correlation of Fig.~\ref{fig:sSFR_logC} indicates that the SED shape alone is able to discriminate other stages during the interaction.

The normalization of the MS declines from $z=2.5$ to $z=0.0$ \citep{Noeske07, Daddi07, Magdis10b, Elbaz11, Magnelli14}, which implies that the vertical dashed line in Fig.~\ref{fig:sSFR_logC_a} moves towards larger values of sSFR at higher redshifts. The relative offset of the intermediate-$z$ LIRGs in the figure could be due to this effect. In order to compare all systems with respect to the location of the $z=0$ MS, in Fig.~\ref{fig:sSFR_logC_b} we have shifted the intermediate-$z$ LIRGS to match the sSFR that they would have if they were in the Local Universe, according to the parametrization of \citet{Elbaz11} (their Eq.~13). We then fit three different power laws ($\log\, \mathcal{C} = a\times \log\, \rm{sSFR} + b$) to the resulting correlation, each corresponding to one of the following subsets: the simulations prior to coalescence (filled circles), the observed local interactions (red triangles)\footnote{Four galaxies have been excluded from the fit: M51b is a post-starburst system where star formation has been quenched \citep{Cooper12}; NGC~3185 and NGC~3190 both belong to the Hickson 44 compact group, and it has been claimed by \citet{Alatalo14} that suppression of star formation is more likely in such groups; finally, NGC~3031 is an isolated disk more likely to be an analog of the empty circles in the lower part of Fig.~\ref{fig:sSFR_logC}.}, and all the LIRGs, both local and at intermediate $z$  (blue diamonds and green squares, plus NGC~3690, the uppermost red triangle in the figure). In Table \ref{tab:linear_fits} we list the parameters obtained for the power laws.

\begin{deluxetable}{ccc}
\tablecolumns{3}
\tablecaption{Linear fits to the $\log\, $sSFR-$\log\, \mathcal{C}$ correlation}

\tablehead{
  \colhead{Group} &
  \colhead{a} &
  \colhead{b}  
}
\startdata

Pre-coalescence sims. & $1.6 \pm 0.1$  & $20.8 \pm 0.1$  \\
L13 galaxies & $1.3 \pm 0.2$  & $17.4 \pm 0.2$  \\
LIRGs & $1.7 \pm 0.3$ & $20.9 \pm 0.2$  \\

\enddata

\label{tab:linear_fits}
\end{deluxetable}

\subsection{The SFR-$M_*$ plane}
\label{sec:MS}

\begin{figure}[!ht]
  \centering
  
\includegraphics[scale=0.36,angle=0,trim=0.0cm 0cm 0.0cm 0.0cm,clip=true]{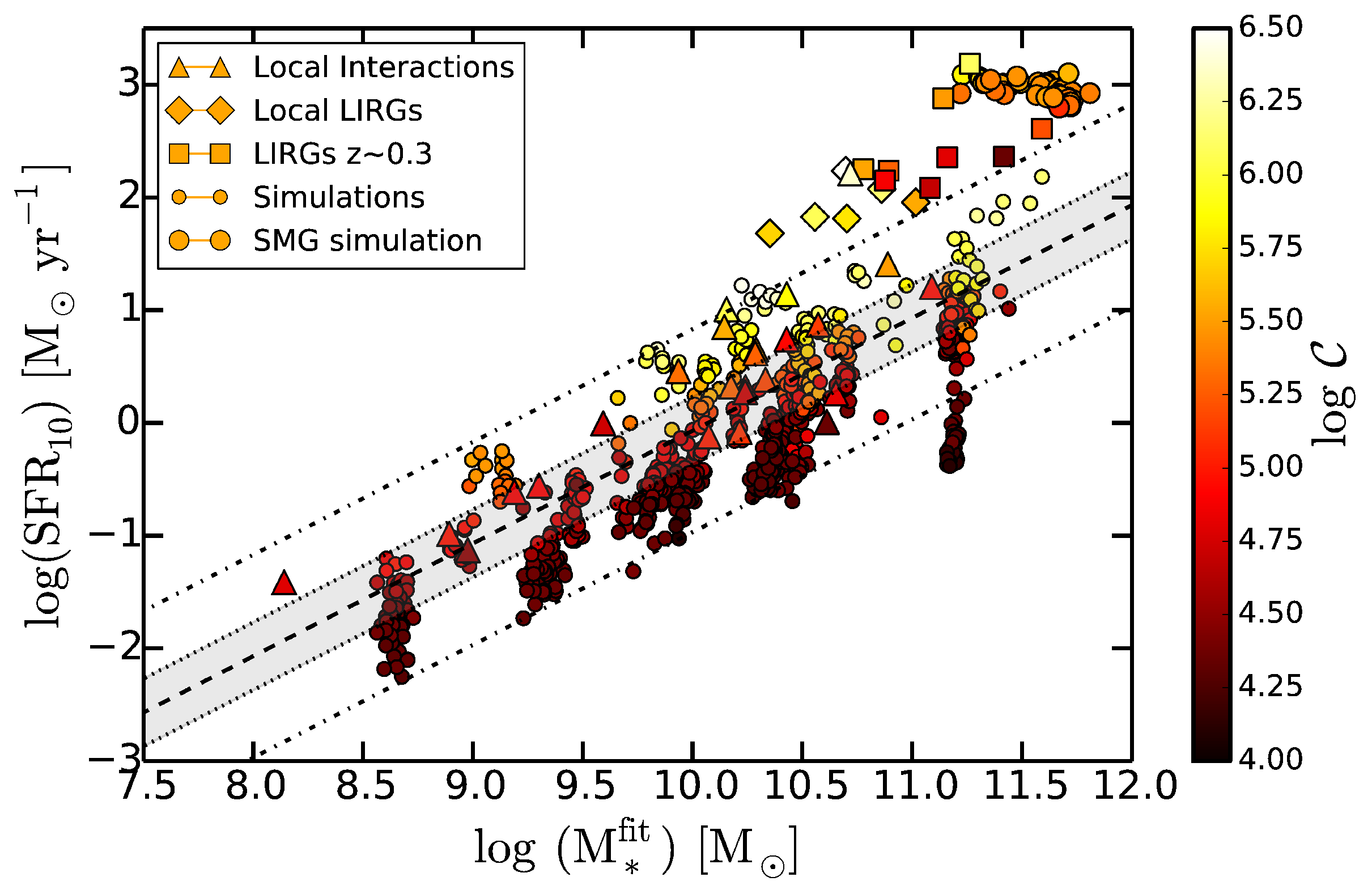}

  \caption{The $\log\, $SFR-$\log\, M_*$ correlation for the simulated and observed systems. Symbols are the same as in Fig.~\ref{fig:sSFR_logC} and are color-coded by compactness. The dashed line is the $z=0$ MS, with 0.3~dex limits indicated by the shaded area. Also shown are the 0.9~dex limits as dot-dashed lines.}

\label{fig:ms_all}
\end{figure}

In Fig.~\ref{fig:ms_all} we plot the simulation snapshots and the observed systems in the SFR-$M_*$ plane, the plane on which the MS correlation is usually defined. The same $z=0$ MS as in Fig.~\ref{fig:sSFR_logC_a} is indicated by the dashed line and the shaded area, and we have added dot-dashed lines to indicate a distance of 0.9~dex from the MS. We observe a gradient in compactness as a function of distance to the MS in the simulations, with compactness increasing from the bottom to the top envelope of the MS. Practically all simulations involving the M0 to M4 progenitors fall within the 0.9~dex limits of the MS, at all times\footnote{A notable exception are the post-coalescence snapshots of M4-M4, which all fall below our 0.9~dex limit.}, but only a fraction of them fall within the 0.3~dex limits. The gas-rich SMG simulation, on the other hand, is a clear outlier of the correlation, sitting clearly above the 0.9~dex limit. The local interactions overlap with the simulations on this plane, whereas both the local and intermediate-$z$ LIRGS are significantly above this correlation.

It is important to note here that stages close to coalescence are over-represented in the sample of snapshots shown here, except for M4-M4, because the time resolution used to generate the SEDs is larger near the peak of infrared luminosity. In other words, in the real world galaxies spend only a small fraction of their lifetimes as strongly disturbed mergers (at high compactness stages), and a similar diagram weighted with a realistic merger fraction would look more crowded in the low compactness end. This means that the scatter shown if Fig.~\ref{fig:ms_all} is not representative of the scatter seen in the observed MS. Nevertheless, the trend shown here is relevant in the context of the evolution of dust geometry with respect to the stars along the merger sequence, and the effect that this evolution has in shaping the MS. Undisturbed disk galaxies are located somewhere near the lower envelope of the correlation, where the majority of real galaxies lie along the MS, and as they evolve towards coalescence they move away from the MS, and to higher values of compactness. In our hydrodynamical models, this increase in compactness translates into a smooth evolution of the large scale geometry of the dust with respect to the stars, driven by higher pressures and/or stronger radiation fields, as galaxy pairs approach coalescence.

For a binary merger to have a very high compactness ($\log~\mathcal{C}>6.0$), its mass ratio has to be very close to 1. Only the most massive mergers reach a very compact stage, and this only during the coalescence phase. However, not even the massive M4-M4 merger can produce the observed outlier sequence of LIRGs. Looking at the correlation between sSFR and compactness for the simulations we observe that as mergers become more massive for a given gas fraction ($f_{\rm{gas}}\sim 0.2-0.3$), they move upwards along the same diagonal correlation in Fig.~\ref{fig:ms_all}. The gas-rich SMG simulation ($f_{\rm{gas}}\sim 0.2-0.6$), on the other hand, is the only one of all our simulations that appears significantly above the main sequence, indicating that only a significant increase in gas fraction can produce the population of LIRG outliers. This is true even if such gas-rich simulations are not yet in the coalescence phase.

\subsection{The IR8 parameter}

We now estimate the parameter $\rm{IR8} = L_{IR}/L_{8\mum }$ defined in \citet{Elbaz11} for our set of simulated and observed interactions. \citet{Elbaz11} show evidence that the vast majority of IR8 values for galaxies in the GOODS-\emph{Herschel} field follow a Gaussian distribution with median $\rm{IR8}\sim 4$. They claim that such distribution defines an infrared Main Sequence. The outliers of this infrared MS ($<20$\%) form a tail towards larger values of IR8 and typically have IR8$\sim 10$. They also show that these outliers are systems with compact projected star formation densities. Here we investigate whether these features of the infrared MS are also present in our simulated and observed systems.

We measure IR8 from the best-fitting SEDs obtained with \textsc{Chiburst}. We integrated the rest-frame SEDs between 8\mum\ and 1000\mum\ to estimate the value of $L_{IR}$ for each system. For observed and simulated systems we then used the (measured or mock) flux density at 8~\mum\ as an estimate of $L_{8}$. For those observed systems for which we did not have measured rest-frame 8\mum\ flux densities available, we integrated the SEDs convolved with the IRAC 8\mum\ filter response, and used the resulting value as $L_8$. We took care in performing the same process for the intermediate-$z$ LIRGs, using their rest-frame SEDs.

\begin{figure}[!ht]
  \centering
  
\includegraphics[scale=0.35,angle=270,trim=0cm 2.0cm 0cm 0cm,clip=true]{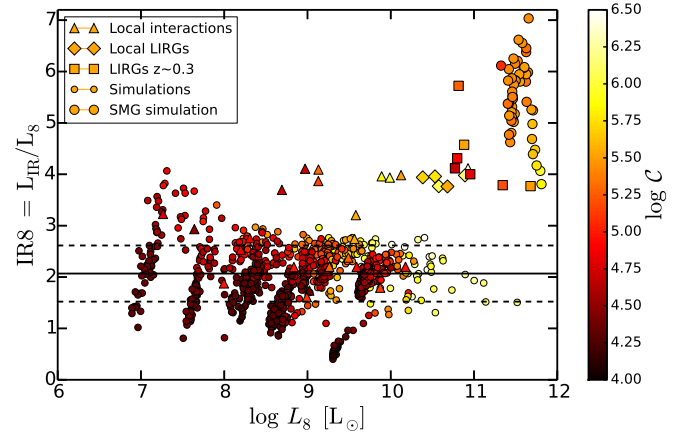}

  \caption{The $\rm{IR8} = L_{IR}/L_{8\mum }$ parameter as a function of 8~$\mum$ luminosity for simulated and observed systems. The symbols are color-coded according to the compactness of the systems. The solid horizontal line represents the mean value of the IR8 values for the simulated systems, excluding the SMG simulation, whereas the dashed lines mark the 1$\sigma$ boundaries around the mean.}

\label{fig:IR8}
\end{figure}

Fig.~\ref{fig:IR8} shows IR8 as a function of the 8~\mum\ luminosity for all simulated and observed systems, color-coded according to their compactness. For the simulated systems (except the SMG simulation), the mean of the distribution is IR8$ = 2$ (solid line in Fig.~\ref{fig:IR8}), and the standard deviation is 0.12~dex. The majority of simulated snapshots are within 1$\sigma$ of the mean value (indicated by the dashed lines), and outliers include systems with both higher and lower values of IR8. Simulated outliers with higher values of IR8 correspond to low luminosity systems ($L_8 < 10^8~\rm{L}_{\odot}$, isolated M0, isolated M1, and M0-M0 interaction). This resembles recent results in a survey of local galaxies by \citet{Cook14}, who find a departure from the infrared MS towards higher IR8 values in low luminosity systems and attribute this to the suppression or destruction of PAH molecules. \textsc{Chiburst} also finds smaller PDR covering fractions( $f_{\rm{PDR}} < 0.4$) in these systems. In this case, however, rather than PAH destruction, the effect is due to a smaller filling factor of PDRs. No clear indication of a high-IR8 tail can be inferred from these simulated interactions. Nevertheless, observed outliers of the MS in Fig.~\ref{fig:ms_all} also appear as outliers of the IR8 distribution.

Objects with high compactness are located at the luminous end of the distribution, and their IR8 values are within the 1$\sigma$ boundaries. About half of the observed local interactions (triangles in Fig.~\ref{fig:IR8}) have higher values of the IR8 parameter (mean is 2.9 and standard deviation of 0.13~dex). The SMG simulation has IR8 values well above the distribution delimited by the dashed lines, with a mean value of IR8 = 5.5. Unlike the low luminosity systems, however, the SMG simulation snapshot SEDs do not show attenuated PAH emission, and therefore in this case we associate the increase in IR8 to an increase in $L_{\rm{IR}}$, which in turn might be associated with the relatively higher fraction of ISM mass in this gas-rich simulation. 

The local and intermediate-$z$ LIRGs also have IR8 values more than 1$\sigma$ above the local interactions and the simulations, but still below the values for the SMG simulation, with a mean IR8 value of 4.1 and a small standard deviation of 0.05. With the limited information we have on the gas content of these systems, it is difficult to assess whether the effect is due to an increases ISM mass, or to suppression/destruction of PAHs. These systems, which are clear outliers of the MS, are also far off the IR8 distribution. Our simulations show that an elevated gas fraction can account for the oddities observed in disks, both in the \emph{regular} MS and the infrared MS.

\begin{figure*}[!t]
  \centering
  \subfigure[]{
\includegraphics[scale=0.38,angle=0,trim=2.0cm 0.1cm 3.0cm -0.1cm,clip=true]{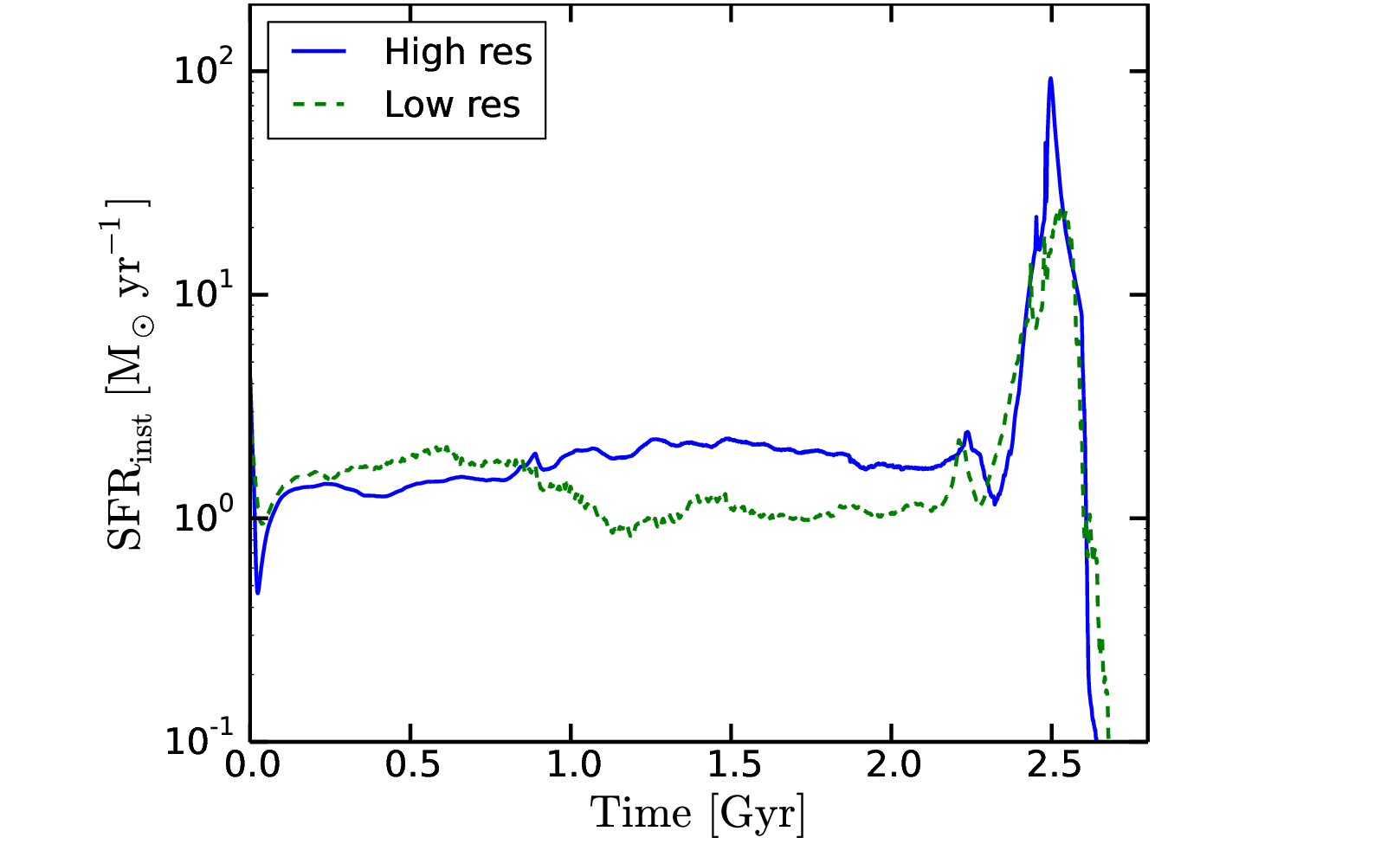}
\label{fig:res_effects1}
}
  \subfigure[]{
\includegraphics[scale=0.38,angle=0,trim=2.0cm 0cm 3.0cm 0cm,clip=true]{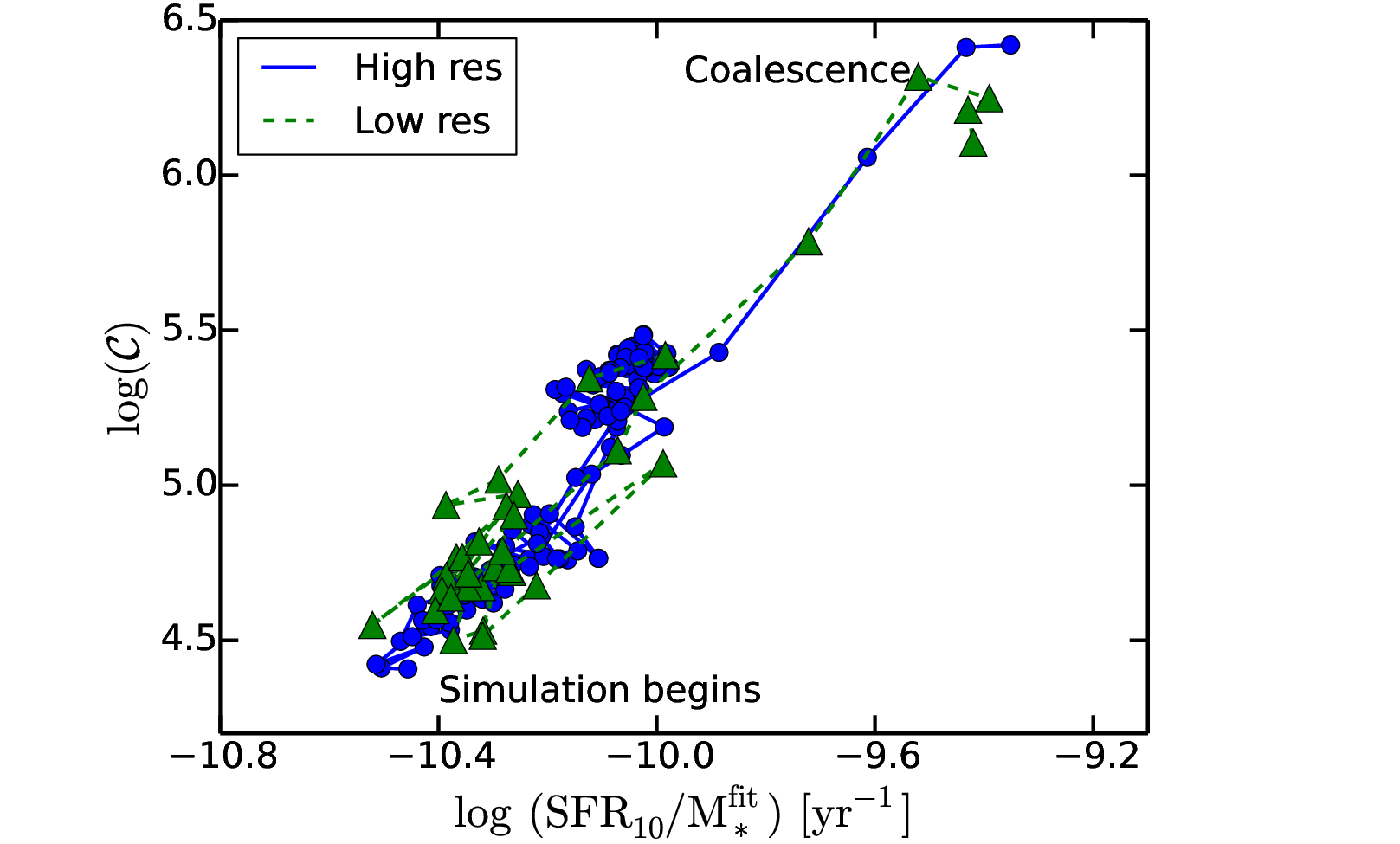}
\label{fig:res_effects2}
}

  \caption{\emph{(a)} The instantaneous SFR as a function of time for the simulated M3-M3 merger, for two different SPH resolutions. The dashed green line represents the standard simulation described in L14, whereas the solid blue line corresponds to the same simulation, but using 8 times more particles and half the smoothing lengths. \emph{(b)} The resulting evolution of the  the sSFR and compactness for the same pair of simulations, as derived using our SED fitting method. The green triangles represent the various snapshots in the standard simulation, whereas the blue dots represent snapshots of the high resolution observation. While the snapshots in each case do not correspond to the same times (we have used a higher time resolution for the high resolution simulation), they cover the same time range between first approach and coalescence.}

\label{fig:true_vs_derived}
\end{figure*}

The mean value of IR8 for the simulated systems in Fig.~\ref{fig:IR8} is a factor of 2 lower than the value reported in \citet{Elbaz11}, for both their local sample of galaxies and the high redshift sample. A number of reasons could explain this disagreement. First, because of the heterogeneous nature of their local sample, both their $L_{IR}$ and $L_8$ were estimated based on the available photometry, and no homogeneous method was used. In particular, for high-$z$ galaxies, they were relying on SEDs that were usually incomplete in the long-wavelength end, since this part of the spectrum had been redshifted beyond the \emph{Herschel} bands. Furthermore, they use a different set of SED templates \citep[namely the][templates]{Chary01} to perform the fits to the galaxy photometry. These templates have a similar treatment of the dust particles as our models, but differ in the treatment of the emission from PAHs. Whereas \textsc{Chiburst} uses an empirical PAH spectrum with optical properties similar to those of coronene, and fitted to two particular interacting galaxies, the \citet{Chary01} templates use a mixture of six different planar PAH, and their emissivities are explicitly calculated.

\subsection{Convergence with respect to spatial resolution}

Convergence of the mock SEDs with respect to the SPH resolution of the \textsc{Gadget-3} simulations is necessary in our analysis. In order to demonstrate that our results do not strongly depend on resolution effects, we have performed an additional version of the simulated M3-M3 merger with 8 times more particles, which implies that the particle mass is 8 times smaller. We have also used softening lengths that are half those of the standard runs. In Fig.~\ref{fig:res_effects1} we show the resulting difference in the SFH due to the increase in resolution. In general, the behavior is very similar for both resolutions, and we only observe minor differences along the duration of the simulation. The largest deviations occur between the first and the second encounter (between 1 and 2~Gyr from the start of the simulation), and at the peak of star formation during coalescence (at $\sim$2.5~Gyr). At these stages, the SFRs can differ by a factor of about 2 between the low and high resolution simulations. These differences do not significantly affect our result in Fig.~\ref{fig:sSFR_logC}, because changes in sSFR due to resolution effects are also accompanied by consistent changes in compactness, as can be seen in Fig.~\ref{fig:res_effects2}, where we show that snapshots from the high resolution simulation overlap with those from the standard run in the $\log\, $sSFR-$\log\, \mathcal{C}$ space. This implies that the correlation derived between sSFR and compactness holds at different SPH resolutions, within the uncertainties indicated. In support of this result, \citet{Hayward14a} show that in the case of mergers up to coalescence (although possibly not after coalescence), the physics derived using SPH methods are robust to the inaccuracies inherent to SPH itself. 

Fig.~\ref{fig:res_effects2} is also useful to visualize how the $\log\, $sSFR-$\log\, \mathcal{C}$ correlation relates to the time evolution of the merger: at initial approach, the simulation starts off with low sSFR and low compactness. As the merger approaches coalescence, there is a boost of the star formation activity, the radiation field intensity in a given volume of gas increases, and gas compression produces an increase in pressure. Therefore, both compactness and sSFR increase until they reach a maximum at coalescence, and in the process, a particular galaxy moves along the correlation in Fig.~\ref{fig:sSFR_logC}. For the gas fractions considered in the simulations, that range from $f_{\rm{gas}}\sim 0.16$ to $f_{\rm{gas}}\sim 0.36$ (see Table \ref{tab:progenitors}), this evolution occurs along the same power law and the same normalization. Only the SMG simulation, which has a relatively larger gas fraction ($f_{\rm{gas}}\sim 0.6$) appears shifted towards larger values of sSFR in this diagram.

\section{Discussion}
\label{sec:discuss}

\subsection{Implications for the Main Sequence}

In intermediate and high redshift surveys of star-forming galaxies, the observed scatter across the MS as well as the existence of a population of outliers pose challenging questions regarding the processes controlling star formation inside galaxies. A paradigm has emerged according to which galaxies that belong to the MS have secular star formation histories maintained by the accretion of gas from the intergalactic medium, whereas outliers of the MS form stars in timescales significantly shorter than their dynamical timescales. Based on morphological studies of a large sample of galaxies with $z$ up to 2.5, \citet{Wuyts11} find that a majority of galaxies in the MS show a disk-like morphology, which supports this scenario.  Several studies, however, have found a significant population of mergers (up to 20\%) on the MS \citep{Kartaltepe12, Hung13}. In this scenario, the intrinsic scatter of the MS arises from the stochasticity in the gas accretion history\footnote{Here stochasticity understood as the result of a Markov chain, when the next state of star formation depends only on the current state.} \citep{Whitaker14, Kelson14}. 

In this context, some relevant questions are: Do the heating conditions of the dust change across this correlation? Are mergers in a stage of coalescence responsible for the population of outliers of the MS? Here we have shown that gas content is not the only possible source of scatter in the MS. Instead, our results (Figs.~\ref{fig:sSFR_logC} and \ref{fig:ms_all}) reveal that, for moderate gas fractions, the large-scale geometry of the dust evolves smoothly across the MS as interacting pairs approach coalescence, from less to more compact, and that this process also moves galaxies away from the MS, towards higher star formation efficiencies. This change in compactness is manifested in the observed sSFR-$T_{\rm{dust}}$ correlation. Evidence for variations in the ISM physics across the MS for a given redshift has been collected by several authors \citep{Daddi10b, Saintonge11, Magnelli14}, although some studies have suggested that the distance from a particular galaxy to the locus of the MS depends only on $f_{\rm{gas}} = M_{\rm{gas}}/M_*$ \citep{Magdis12}.

Mergers therefore result in mobility of galaxies across the MS, but because of the short duration of the coalescence phase, the sharp increase of star formation during coalescence has only a minor impact on the scatter of the observed MS. Also, extreme outliers of the MS such as LIRGs fall significantly above the sequence, regardless of the mass of the interacting galaxies. This implies that other mechanisms should be invoked in order to explain their extremely high sSFR. One possibility is that they have significantly higher gas fractions ($f_{\rm{gas}}$), as our SMG simulation demonstrates. The fact that LIRGs are also outliers of the infrared MS suggests that the change in physical conditions associated with the jump to more efficient star formation also makes the ratio of total infrared luminosity to PAH emission weaker. Our results indicate the this is not always associated with suppression of the PAH emission.

\subsection{Relevance of compactness: difference with $T_{\rm{dust}}$.}
\label{sec:comp_vs_T}

A fundamental difference between compactness and $T_{\rm{dust}}$ relies on the fact that, whereas $\mathcal{C}$ is a \textit{parametrization} of the compactness of star formation -that is, a measurement of the geometry of the star-forming ISM with respect to the (ionizing) stars-, $T_{\rm{dust}}$ is a \textit{result} of this geometry (and of other factors affecting the dust heating in galaxies, as discussed in \citet{Groves12}). 

\begin{figure}[!ht]
  \centering
  
  \includegraphics[scale=0.38,angle=270,trim=0cm 1.8cm 0cm 0cm,clip=true]{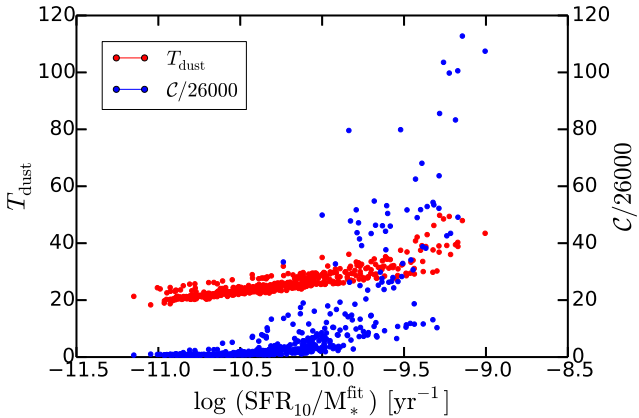}

  \caption{The sensitivity of $T_{\rm{dust}}$ and $\mathcal{C}$ to changes in sSFR. The plotted compactness and sSFR are from the fits to the simulations described in this paper, whereas the $T_{\rm{dust}}$ data are from \citet{Lanz14}. Compactness has been renormalized by dividing by a factor of 26000 to make the comparison more evident.}

\label{fig:comp_Tdust_logC}
\end{figure}

In the \citet{Groves08} models, compactness parametrizes the time evolution of $T_{\rm{dust}}$ as a function of the ISM radiation field and pressure. In the context of this paper, changes in compactness are related to variations in the distribution of dust temperatures as a function of sSFR. Even though the two quantities $\mathcal{C}$ and $T_{\rm{dust}}$ are closely related, they do not necessarily provide the same information. For example, $T_{\rm{dust}}$ can be influenced by heating sources other than the young massive stars in clusters \citep[see, for example][]{Groves12} and by variation in the radiation field (i.e. young vs. old bursts), whereas compactness parametrizes purely the geometry of the dust associated to star formation. As an example of the differences between $\mathcal{C}$ and $T_{\rm{dust}}$, we note that in the high sSFR end ($\log\, \rm{sSFR} > -10~\rm{yr}^{-1}$), the compactness parameter is significantly more sensitive to changes in the sSFR than the $T_{\rm{dust}}$, as shown in Fig.~\ref{fig:comp_Tdust_logC}. Stages with such large values of sSFR correspond in our simulations to the coalescence phase of the most massive interactions, when the star formation activity peaks (and presumably, when the ISM conditions are changing more rapidly and the bolometric luminosity has the highest contribution from star-forming ISM). By using compactness, we can therefore characterize the most intense epochs of star formation as a function of the ISM properties, something that it is not possible based on $T_{\rm{dust}}$ measurements only.

\subsection{Global compactness and the physical size of star-forming galaxies.}

The large scale ($>200$~pc) geometrical structure of the star-forming ISM is reflected in the compactness values that we have estimated for the simulated interactions. At these scales, a larger value of $\mathcal{C}$ implies a higher projected star formation density, because both the average radiation field and the gas density increase in areas of concentrated star formation. Therefore, the properties of the correlation found between $\mathcal{C}$ and sSFR (Fig.~\ref{fig:sSFR_logC} and Table \ref{tab:linear_fits}) contain information regarding the large scale geometry of star-forming gas as a function of the total star formation taking place in these systems. We use the slope of the correlation to quantify this relation. 

We start with the fundamental definition of compactness in Eq.~\ref{eq:logC_original}, and assume that the proportionality holds for all times. The amount of luminosity being produced in a patch of ISM with radius $R$ scales linearly with the sSFR within that particular patch, and hence:

\begin{equation}
\label{eq:logC_uno}
\log \mathcal{C}=\log\, \rm{sSFR} - 2\log\, R + constant.
\end{equation}

Now, we have already estimated the slope of the correlation to be $\sim 1.6$. Therefore we have:

 \begin{equation}
\label{eq:logC_dos}
\log \mathcal{C}=1.6~\log\, \rm{sSFR} + constant.
\end{equation}

If we compare Eqs. \ref{eq:logC_uno}  and \ref{eq:logC_dos}, then we can infer that if $\rm{sSFR}^{-0.3}\propto R$, then we recover the measured slope of 1.6 for the $\log\, $sSFR-$\log\, \mathcal{C}$ correlation. A possible interpretation for this is that systems with high sSFR have smaller physical sizes. For a fixed gas mass, the volumetric S-K relation implies a similar effect, i.e., that the sSFR should be inversely proportional to the volume of the region in consideration. Qualitatively, this is also consistent with observations of local LIRGs. For example, approximately half of the luminosity of Arp~220 is emitted from within its central 100~pc, which means that this object with very high sSFR is also very compact.

\subsection{Small-scale effects}

The observed segregation between normal galaxies and LIRGs in (Fig.~\ref{fig:ms_all}), resembles the bimodality found by some authors in the S-K law, that relates the projected molecular gas density with the projected star formation density. For example, \citet{Daddi10b} find that disk galaxies and starbursts occupy different regions in the gas mass versus SFR plane. They argue that the existence of these two regimes is related to different fractions of molecular gas in the ISM, and that only accounting for the dynamical timescales of galaxies recovers a single, universal star formation law. A different approach is adopted in \citet{Krumholz12}, where the authors claim that the observed variations are due to geometrical effects, and that all objects, from small molecular clouds to LIRGs, follow a universal volumetric star formation controlled by the local free fall times. 

In principle, we could explore the implications of our results (in particular the measured slopes in Table \ref{tab:linear_fits}) under the assumption that the \citet{Krumholz12} interpretation holds, and attempt to constrain the physics of the small scale star-forming ISM, such as pressure and average cluster mass. However, this would only be possible if the simulations used are fine enough to resolve such smaller scales ($<100$~pc). Given our resolution limitations, we leave this endeavor for an upcoming paper.

\medskip
\medskip
\medskip

\subsection{The relation between star formation and gas content in local mergers and LIRGs}
\label{sec:offsets}
The offset between LIRGs and local interactions in Figs.~\ref{fig:sSFR_logC} and \ref{fig:ms_all} implies one or several of the following possibilities: they have different gas fractions ($f_{\rm{gas}}$), they have different star formation efficiencies, or they have different conditions in pressure and radiation field that result in a different compactness for a given bolometric luminosity. Enhanced gas fractions can rise galaxies significantly above the MS, even in early quiescent, disk-like stages, as demonstrated by our results on the SMG simulation. We examine here whether an increased gas content in the observed LIRGs is responsible for their increased sSFR.

As we have mentioned, the normalization of the MS declines from $z=2.5$ to $z=0.0$, and the reasons for this decline are not entirely understood. Recent \emph{ALMA} observations suggests that all galaxies at all cosmic times convert gas into stars with the same efficiency, but that galaxies at intermediate and high $z$ have relatively larger gas reservoirs \citep{Scoville15}. On the other hand, using a multi-wavelength stacking analysis of the SEDs of galaxies in the COSMOS field, \citet{Bethermin15} show that even if the bulk of star formation up to $z\sim 4$ is dominated by secular processes, star formation efficiencies in strong starburst differ from those in MS galaxies.

Using the molecular gas mass estimates for the intermediate-$z$ LIRGs in \citet{Magdis14}, which are based on SED fitting and an assumed gas-to-dust ratio, we find that these galaxies have typical $f_{\rm{gas}}$ of about 0.7-0.8 (and as high as 1.1 for SWIRE7, which brings it close to a typical submillimeter galaxy). This is about a factor of 3 higher than the value of $f_{\rm{gas}}$ in the simulations and local mergers, including the luminous stage 4 mergers \citep{Sanders91}. This suggests that, in agreement with our SM simulation, gas content plays an important role in creating the population of outlier LIRGs in the MS.

Finally, we want to emphasize that regardless of the normalization factor, the slope of the  $\log\, $sSFR-$\log\mathcal{C}$ is very similar for simulations, normal galaxies, and LIRGs. This suggests that LIRGs might be gas-rich analogs of local galaxies, perhaps undergoing similar merging events. Confirming this would require morphological information that we currently lack for the intermediate-$z$ LIRGs, but it is worth investigating it. Nevertheless, the large range of compactness values that these luminous systems span indicates that not all LIRGs have compact star formation, and that at least some of them should have disk-like morphologies. In fact, \citet{Magdis14} show that the $z\sim 0.3$ LIRGs have $L_{[\cii\ ]}/L_{\rm{IR}}$ ratios similar to those of local disks. Regarding the morphology of LIRGs and ULIRGs, recent morphological and kinematical studies have revealed an increasing number of non-interacting, disk-dominated ULIRG-like systems with look-back time \citep[e.g.][]{Forster_Schreiber09, Kartaltepe12, Wang12}. Larson et al. (in preparation ) also find evidence for non-interacting morphology well above the MS in a sample of nearby galaxies.

\subsection{Limitations of the hydrodynamical models}
\label{sec:limitations}
The hydrodynamical simulations used in this paper employ a simplified model for the ISM. First, with respect to other recent simulations, they assume a relatively low density threshold for the onset of star formation. Also, they employ a pressurized equation of state, which might have the effect of artificially smoothing the ISM, thus affecting the emerging far-IR emission\footnote{However, in \citet{Hayward14a}, it is demonstrated that the phase structure of the gas is very insensitive to the numerical scheme employed, except for the hot halo gas (which is unimportant for the radiative transfer). This can be seen very clearly in the animations available here (https://www.cfa.harvard.edu/itc/research/arepomerger/) that show the evolution of the gas phase diagrams. Given the striking similarity in the phase diagrams, especially in the dense gas, it is unlikely that the differences that arise from different hydro techniques will lead to significant differences in the radiative transfer results, and hence in the SED-derived parameters.}. Other groups have used higher density thresholds and more sophisticated models for the ISM. Such simulations certainly feature more complicated small-scale ISM structure and might be able to more accurately reproduce its conditions at scales smaller that the resolution limit of the simulations presented here, which is of the order of a few hundred parsecs. This limitation in ISM physics notwithstanding, the usefulness of our approach lies on our ability to parametrize the large scale ($>200-300$~pc) structure of the ISM and to study how this relates to the star formation properties at even larger scales (e.g., the global compactness of galaxies as they evolve across the MS and the global SFR). Furthermore, we note that even though we do not resolve the small-scale structure of the ISM, the trends in star formation properties that we derive in \S~\ref{sec:compact_correl} and \S~\ref{sec:MS} for the simulations are in good agreement with what is observed in real interacting galaxies. This direct comparison of simulated and real galaxies through their parametrized SEDs therefore implies that the small scale physics that control star formation in galaxies are imprinted in the ISM structure at larger scales, and that it is currently possible to study such structure.

 \section{Summary}
\label{sec:summary}

The majority of star-forming galaxies follow a simple empirical correlation in the SFR versus stellar mass ($M_*$) plane, of the form $\mbox{SFR} \propto M_*^{\alpha}$. The physics behind the MS is currently a subject of debate, and no consensus has been reached regarding the local conditions of the gas and the dust that set the basic properties of the MS. In order to advance our understanding of the nature of the MS, we have combined a set of hydro-dynamical simulations of isolated disk galaxies and binary mergers with state-of-the-art radiative transfer codes to analyze the physics of the star-forming ISM along the interaction sequence. Using the SED-fitting code \textsc{Chiburst}, we have fitted the SEDs of simulated interactions, local interacting galaxies and intermediate-$z$ LIRGs and derived their star-forming properties. In particular, we parametrize the ISM conditions at scales larger than a few hundred pc via the compactness parameter ($\mathcal{C}$), which controls the distribution of dust temperatures as a function of time. We have reached the following conclusions:

\begin{enumerate}

\item In addition to SFR and $M_*$, compactness ($\mathcal{C}$) is a useful parameter  to describe the properties of galaxies along and across the main sequence. Variations in $\mathcal{C}$ as a function of sSFR underlie the observed variations in dust temperatures with infrared luminosity in star-forming galaxies. Whereas $\mathcal{C}$ is a \textit{parametrization} of the compactness of star formation (that is, a measurement of the geometry of the star-forming ISM with respect to the ionizing stars), $T_{\rm{dust}}$ is a \textit{result} of this geometry.

\item During the course of a galactic merger, as the interacting pair approaches coalescence, the geometry of the ISM in scales larger than a few hundred parsecs evolves from a less compact to a more compact state. This results in the sSFR-$T_{\rm{dust}}$ correlation observed in several galaxy surveys. Here we have derived a more fundamental correlation between sSFR and compactness, of the form $\mathcal{C} \propto \rm{sSFR}^{1.6\pm 0.1}$.

\item As this evolution from less to more compact geometry takes place, the depletion of gas to fuel star formation becomes more efficient and galaxies also depart from the MS at coalescence (with the sSFR increasing up to about 1 order of magnitude), contributing to the scatter in the sequence. However, because mergers spend very little time in a coalescence stage, it is unlikely that mergers alone can explain the observed scatter of the MS.

\item LIRGs fall significantly above ($> 1.0~\rm{dex}$) the MS correlation. In order to reproduce these extreme outliers, significantly higher gas fractions ($f_{\rm{gas}}\sim 0.6$) are required, as in our SMG-type simulation. We have shown that, in fact, our sample of intermediate-$z$ LIRGs have enhanced gas fractions with respect to local mergers.

\item LIRGs are also well above the infrared MS defined based on the $\rm{IR8} \equiv L_{\rm{IR}}/L_8$ diagnostic. This suggests that the increase in gas fraction is also related to a higher ratio of total infrared luminosity to PAH emission. Our results indicate the this relative decrease of intensity of the PAH bands is not necessarily associated with suppression of the PAH emission, but rather to a relative enhancement of the total infrared emission.

\item Systems with large sSFR have relatively small physical sizes, according to: $R \propto \rm{sSFR}^{-0.3}$. For a fixed gas mass, the S-K relation implies a similar trend, that is compatible with observations of luminous, compact  starbursts like Arp~220.

\item The observed segregation between normal galaxies and LIRGs in the $\log\, \mbox{sSFR}$-$\log\, \mathcal{C}$ plot conserves the slope of the correlation regardless of normalization. This suggests that LIRGs might be gas-rich analogs of local galaxies, perhaps undergoing similar merging events. Confirming this would require morphological information that we currently lack for the intermediate-$z$ LIRGs, but it is worth investigating if a fraction of intermediate-$z$ LIRGs have disk-like morphology that correlates with low compactness.

\end{enumerate} 

\acknowledgements 
We thank the anonymous referee for very useful remarks on the paper. The authors would also like to thank Dimitra Rigopoulou and Georgios Magdis for providing the photometry for the $z\sim 0.3$ ULIRGs. JRMG, HAS and LL acknowledge partial support from NASA grants NNX10AD68G and NNX14AJ61G, and JPL RSA contracts 1369565 and 1369566. This research has made use of NASA’s Astrophysics Data System Bibliographic Services. The simulations in this paper were performed on the Odyssey cluster supported by the FAS Research Computing Group at Harvard University.

\begin{appendix}
\label{app:fitting}

We have interpolated from the original grid of galaxy SED models to create a continuous parameter space for the star forming parameters described above. Additionally, dust extinction due to diffuse dust in the galactic systems studied (i.e., not associated with \hii\ regions) is also a free parameter in our models, parametrized using the visual extinction in magnitudes ($A_V$). We use the attenuation law of \citet{Fischera05} that approximates the empirical Calzetti extinction law for starburst galaxies \citep{Calzetti01}. With the resulting set of model parameters, we attempt to fit the observed SEDs of galaxies.

\textsc{Chiburst} is a Bayesian Monte Carlo fitting routine to fit multi-wavelength observations of galaxies using the models described above. Given a set of photometric or spectroscopic data and their respective uncertainties, \textsc{Chiburst} uses Bayesian inference to calculate posterior PDFs for the following model parameters: the stellar mass ($M_*$), SFR$_{10}$, SFR$_{100}$, SFR$_{1}$, the PDR covering fraction ($f_{\rm{PDR}}$), the compactness parameter ($\mathcal{C}$), the ISM ambient pressure ($P/k$), the mean metallicity of the system (Z), and the visual extinction in the line of sight towards the system ($A_V$).

Here is how it works: starting from Bayes's theorem, \textsc{Chiburst} calculates posterior PDFs for the model parameters as the product of two distributions: the likelihood that the data can be drawn from a particular combination of model parameters, and a prior distribution for the model parameters that accounts for any \textit{a priori} beliefs that we may have regarding the possible values for the parameters. For each parameter, the likelihood is obtained from the distribution of $\chi^2$ values resulting from comparing the dataset with the models. The obtained posterior PDFs are the most complete solution that we can obtain given the available data, the parameter space of models, and our previous believe about the parameter values.

\subsubsection{The Probability Distribution Functions}
Suppose that you have obtained photometry of a galaxy in different bands, with certain observational uncertainties associated. Bayes’ Theorem states that, given those observations, the probability $P(M|D)$ of an SED model $M$ being a true representation of the observed galaxy SED data $D$, is proportional to the product of the likelihood that your data-points can be obtained from your model ($P(D|M)$) times a prior distribution $P(M)$ that contains independent evidence of certain model parameters having certain values. $P(M|D)$ is what we call the posterior PDF and is the solution we are after. The likelihood $P(D|M)$, or the probability of the data given the model, can be obtained from the distribution of reduced $\chi^2$ values if we assume that the observational errors are Gaussian, i.e., if we assume that multiple measurements of the flux at a particular band will distribute according to a Gaussian. The expression for the likelihood is then:

\begin{equation}
\label{eq:likelihood}
P(D|M) = \sum_i \exp{(-1/2\, \chi_{\rm{red},i}^2)} ,
\end{equation}

\noindent
where the sum is marginalized for each model parameter over all possible $\chi^2_{\rm{red},i}$ values for models with a given value of the parameter. The \textit{prior} $P(M)$ is a measure of any previous knowledge that you have on a particular parameter or set of parameters. For example, if previous independent measurements indicate that the stellar mass of a galaxy must be within certain values, then you can constrain the possible solutions to your fitting problem by constructing a prior on ($M_{\rm{cl}}$) that is compatible with those values. In the present work we use uniform priors to specify ranges of reasonable value for our parameters, and to bias our posterior PDFs as little as possible. Finally, you need to apply a normalization factor to your posterior PDF to guarantee that the probability of at least one model being a representation for your galaxy equals one.

\subsubsection{Stepping across the parameter space}

Given the infinite size of the parameter space (we interpolate from the original grid to allow any value of the parameters), it is not possible to calculate the posterior PDF for every single allowed value of the model parameters, especially as more data points are added and additional model parameters are considered. Instead, we use a Monte Carlo Markov Chain (MCMC) approach to step across the parameter space while properly sampling the posterior PDF. The idea is simple and is based on the Metropolis-Hastings algorithm: we start at a given location of the parameter space where we can calculate the value of $P(M|D)_{\rm{old}}$ and then randomly move to another location where we calculate the new value of $P(M|D)_{\rm{new}}$. We then calculate the ratio $P(M|D)_{\rm{new}}/P(M| D)_{\rm{old}}$ and compare it to a random number between 0 and 1. If the latter is lower than the calculated ratio, we accept the step and update the value of the model parameters to those of $P(M| D)_{\rm{new}}$. If, on the other hand, the ratio of probabilities is lower than the random number, we reject the step and the model remains unchanged. If the step size and the number of iterations are properly chosen, this process should converge to the posterior PDF. In other words, the histogram of models selected in this fashion should be a representation of the initially unknown posterior PDF, and tells us where in the parameter space are the most likely solutions located, given all the information at hand. This is the ultimate solution to our fitting problem.

\end{appendix}

\nocite{*}

\bibliographystyle{apj}

\bibliography{bib_paper}

\begin{thebibliography}{}
\expandafter\ifx\csname natexlab\endcsname\relax\def\natexlab#1{#1}\fi

\bibitem[{{Adamo} {et~al.}(2011){Adamo}, {{\"O}stlin}, \&
  {Zackrisson}}]{Adamo11}
{Adamo}, A., {{\"O}stlin}, G., \& {Zackrisson}, E. 2011, \mnras, 417, 1904

\bibitem[{{Alatalo} {et~al.}(2014){Alatalo}, {Appleton}, {Lisenfeld},
  {Bitsakis}, {Guillard}, {Charmandaris}, {Cluver}, {Dopita}, {Freeland},
  {Jarrett}, {Kewley}, {Ogle}, {Rasmussen}, {Rich}, {Verdes-Montenegro}, {Xu},
  \& {Yun}}]{Alatalo14}
{Alatalo}, K., {Appleton}, P.~N., {Lisenfeld}, U., {et~al.} 2014, \apj, 795,
  159

\bibitem[{{Armus} {et~al.}(2009){Armus}, {Mazzarella}, {Evans}, {Surace},
  {Sanders}, {Iwasawa}, {Frayer}, {Howell}, {Chan}, {Petric}, {Vavilkin},
  {Kim}, {Haan}, {Inami}, {Murphy}, {Appleton}, {Barnes}, {Bothun}, {Bridge},
  {Charmandaris}, {Jensen}, {Kewley}, {Lord}, {Madore}, {Marshall},
  {Melbourne}, {Rich}, {Satyapal}, {Schulz}, {Spoon}, {Sturm}, {U}, {Veilleux},
  \& {Xu}}]{Armus09}
{Armus}, L., {Mazzarella}, J.~M., {Evans}, A.~S., {et~al.} 2009, \pasp, 121,
  559

\bibitem[{{Behroozi} {et~al.}(2013){Behroozi}, {Wechsler}, \&
  {Conroy}}]{Behroozi13}
{Behroozi}, P.~S., {Wechsler}, R.~H., \& {Conroy}, C. 2013, \apj, 770, 57

\bibitem[{{B{\'e}thermin} {et~al.}(2015){B{\'e}thermin}, {Daddi}, {Magdis},
  {Lagos}, {Sargent}, {Albrecht}, {Aussel}, {Bertoldi}, {Buat}, {Galametz},
  {Heinis}, {Ilbert}, {Karim}, {Koekemoer}, {Lacey}, {Le Floc'h}, {Navarrete},
  {Pannella}, {Schreiber}, {Smol{\v c}i{\'c}}, {Symeonidis}, \&
  {Viero}}]{Bethermin15}
{B{\'e}thermin}, M., {Daddi}, E., {Magdis}, G., {et~al.} 2015, \aap, 573, A113

\bibitem[{{Brassington} {et~al.}(2015){Brassington}, {Zezas}, {Ashby}, {Lanz},
  {Smith}, {Willner}, \& {Klein}}]{Brassington15}
{Brassington}, N.~J., {Zezas}, A., {Ashby}, M.~L.~N., {et~al.} 2015, \apjs,
  218, 6

\bibitem[{{Brinchmann} {et~al.}(2004){Brinchmann}, {Charlot}, {White},
  {Tremonti}, {Kauffmann}, {Heckman}, \& {Brinkmann}}]{Brinchmann04}
{Brinchmann}, J., {Charlot}, S., {White}, S.~D.~M., {et~al.} 2004, \mnras, 351,
  1151

\bibitem[{{Calzetti}(2001)}]{Calzetti01}
{Calzetti}, D. 2001, \pasp, 113, 1449

\bibitem[{{Calzetti} {et~al.}(2010){Calzetti}, {Wu}, {Hong}, {Kennicutt},
  {Lee}, {Dale}, {Engelbracht}, {van Zee}, {Draine}, {Hao}, {Gordon},
  {Moustakas}, {Murphy}, {Regan}, {Begum}, {Block}, {Dalcanton}, {Funes}, {Gil
  de Paz}, {Johnson}, {Sakai}, {Skillman}, {Walter}, {Weisz}, {Williams}, \&
  {Wu}}]{Calzetti10}
{Calzetti}, D., {Wu}, S.-Y., {Hong}, S., {et~al.} 2010, \apj, 714, 1256

\bibitem[{{Castor} {et~al.}(1975){Castor}, {McCray}, \& {Weaver}}]{Castor75}
{Castor}, J., {McCray}, R., \& {Weaver}, R. 1975, \apjl, 200, L107

\bibitem[{{Ceverino} {et~al.}(2010){Ceverino}, {Dekel}, \&
  {Bournaud}}]{Ceverino10}
{Ceverino}, D., {Dekel}, A., \& {Bournaud}, F. 2010, \mnras, 404, 2151

\bibitem[{{Chary} \& {Elbaz}(2001)}]{Chary01}
{Chary}, R., \& {Elbaz}, D. 2001, \apj, 556, 562

\bibitem[{{Ciesla} {et~al.}(2015){Ciesla}, {Charmandaris}, {Georgakakis},
  {Bernhard}, {Mitchell}, {Buat}, {Elbaz}, {LeFloc'h}, {Lacey}, {Magdis}, \&
  {Xilouris}}]{Ciesla15}
{Ciesla}, L., {Charmandaris}, V., {Georgakakis}, A., {et~al.} 2015, \aap, 576,
  A10

\bibitem[{{Cook} {et~al.}(2014){Cook}, {Dale}, {Johnson}, {Van Zee}, {Lee},
  {Kennicutt}, {Calzetti}, {Staudaher}, \& {Engelbracht}}]{Cook14}
{Cook}, D.~O., {Dale}, D.~A., {Johnson}, B.~D., {et~al.} 2014, \mnras, 445, 899

\bibitem[{{Cox} {et~al.}(2008){Cox}, {Jonsson}, {Somerville}, {Primack}, \&
  {Dekel}}]{Cox08}
{Cox}, T.~J., {Jonsson}, P., {Somerville}, R.~S., {Primack}, J.~R., \& {Dekel},
  A. 2008, \mnras, 384, 386

\bibitem[{{Daddi} {et~al.}(2007){Daddi}, {Dickinson}, {Morrison}, {Chary},
  {Cimatti}, {Elbaz}, {Frayer}, {Renzini}, {Pope}, {Alexander}, {Bauer},
  {Giavalisco}, {Huynh}, {Kurk}, \& {Mignoli}}]{Daddi07}
{Daddi}, E., {Dickinson}, M., {Morrison}, G., {et~al.} 2007, \apj, 670, 156

\bibitem[{{Daddi} {et~al.}(2009){Daddi}, {Dannerbauer}, {Stern}, {Dickinson},
  {Morrison}, {Elbaz}, {Giavalisco}, {Mancini}, {Pope}, \& {Spinrad}}]{Daddi09}
{Daddi}, E., {Dannerbauer}, H., {Stern}, D., {et~al.} 2009, \apj, 694, 1517

\bibitem[{{Daddi} {et~al.}(2010{\natexlab{a}}){Daddi}, {Elbaz}, {Walter},
  {Bournaud}, {Salmi}, {Carilli}, {Dannerbauer}, {Dickinson}, {Monaco}, \&
  {Riechers}}]{Daddi10b}
{Daddi}, E., {Elbaz}, D., {Walter}, F., {et~al.} 2010{\natexlab{a}}, \apjl,
  714, L118

\bibitem[{{Daddi} {et~al.}(2010{\natexlab{b}}){Daddi}, {Bournaud}, {Walter},
  {Dannerbauer}, {Carilli}, {Dickinson}, {Elbaz}, {Morrison}, {Riechers},
  {Onodera}, {Salmi}, {Krips}, \& {Stern}}]{Daddi10a}
{Daddi}, E., {Bournaud}, F., {Walter}, F., {et~al.} 2010{\natexlab{b}}, \apj,
  713, 686

\bibitem[{{Dav{\'e}} {et~al.}(2012){Dav{\'e}}, {Finlator}, \&
  {Oppenheimer}}]{Dave12}
{Dav{\'e}}, R., {Finlator}, K., \& {Oppenheimer}, B.~D. 2012, \mnras, 421, 98

\bibitem[{{Dav{\'e}} {et~al.}(2011){Dav{\'e}}, {Oppenheimer}, \&
  {Finlator}}]{Dave11}
{Dav{\'e}}, R., {Oppenheimer}, B.~D., \& {Finlator}, K. 2011, \mnras, 415, 11

\bibitem[{{Dekel} {et~al.}(2013){Dekel}, {Zolotov}, {Tweed}, {Cacciato},
  {Ceverino}, \& {Primack}}]{Dekel13}
{Dekel}, A., {Zolotov}, A., {Tweed}, D., {et~al.} 2013, \mnras, 435, 999

\bibitem[{{Dopita} {et~al.}(2005){Dopita}, {Groves}, {Fischera}, {Sutherland},
  {Tuffs}, {Popescu}, {Kewley}, {Reuland}, \& {Leitherer}}]{Dopita05}
{Dopita}, M.~A., {Groves}, B.~A., {Fischera}, J., {et~al.} 2005, \apj, 619, 755

\bibitem[{{Dopita} {et~al.}(2006{\natexlab{a}}){Dopita}, {Fischera},
  {Sutherland}, {Kewley}, {Tuffs}, {Popescu}, {van Breugel}, {Groves}, \&
  {Leitherer}}]{Dopita06a}
{Dopita}, M.~A., {Fischera}, J., {Sutherland}, R.~S., {et~al.}
  2006{\natexlab{a}}, \apj, 647, 244

\bibitem[{{Dopita} {et~al.}(2006{\natexlab{b}}){Dopita}, {Fischera},
  {Sutherland}, {Kewley}, {Leitherer}, {Tuffs}, {Popescu}, {van Breugel}, \&
  {Groves}}]{Dopita06b}
---. 2006{\natexlab{b}}, \apjs, 167, 177

\bibitem[{{Draine} \& {Li}(2007)}]{Draine07}
{Draine}, B.~T., \& {Li}, A. 2007, \apj, 657, 810

\bibitem[{{Dutton} {et~al.}(2010){Dutton}, {van den Bosch}, \&
  {Dekel}}]{Dutton10}
{Dutton}, A.~A., {van den Bosch}, F.~C., \& {Dekel}, A. 2010, \mnras, 405, 1690

\bibitem[{{Elbaz} {et~al.}(2007){Elbaz}, {Daddi}, {Le Borgne}, {Dickinson},
  {Alexander}, {Chary}, {Starck}, {Brandt}, {Kitzbichler}, {MacDonald},
  {Nonino}, {Popesso}, {Stern}, \& {Vanzella}}]{Elbaz07}
{Elbaz}, D., {Daddi}, E., {Le Borgne}, D., {et~al.} 2007, \aap, 468, 33

\bibitem[{{Elbaz} {et~al.}(2011){Elbaz}, {Dickinson}, {Hwang},
  {D{\'{\i}}az-Santos}, {Magdis}, {Magnelli}, {Le Borgne}, {Galliano},
  {Pannella}, {Chanial}, {Armus}, {Charmandaris}, {Daddi}, {Aussel}, {Popesso},
  {Kartaltepe}, {Altieri}, {Valtchanov}, {Coia}, {Dannerbauer}, {Dasyra},
  {Leiton}, {Mazzarella}, {Alexander}, {Buat}, {Burgarella}, {Chary}, {Gilli},
  {Ivison}, {Juneau}, {Le Floc'h}, {Lutz}, {Morrison}, {Mullaney}, {Murphy},
  {Pope}, {Scott}, {Brodwin}, {Calzetti}, {Cesarsky}, {Charlot}, {Dole},
  {Eisenhardt}, {Ferguson}, {F{\"o}rster Schreiber}, {Frayer}, {Giavalisco},
  {Huynh}, {Koekemoer}, {Papovich}, {Reddy}, {Surace}, {Teplitz}, {Yun}, \&
  {Wilson}}]{Elbaz11}
{Elbaz}, D., {Dickinson}, M., {Hwang}, H.~S., {et~al.} 2011, \aap, 533, A119

\bibitem[{{Fischera} \& {Dopita}(2005)}]{Fischera05}
{Fischera}, J., \& {Dopita}, M. 2005, \apj, 619, 340

\bibitem[{{F{\"o}rster Schreiber} {et~al.}(2009){F{\"o}rster Schreiber},
  {Genzel}, {Bouch{\'e}}, {Cresci}, {Davies}, {Buschkamp}, {Shapiro},
  {Tacconi}, {Hicks}, {Genel}, {Shapley}, {Erb}, {Steidel}, {Lutz},
  {Eisenhauer}, {Gillessen}, {Sternberg}, {Renzini}, {Cimatti}, {Daddi},
  {Kurk}, {Lilly}, {Kong}, {Lehnert}, {Nesvadba}, {Verma}, {McCracken},
  {Arimoto}, {Mignoli}, \& {Onodera}}]{Forster_Schreiber09}
{F{\"o}rster Schreiber}, N.~M., {Genzel}, R., {Bouch{\'e}}, N., {et~al.} 2009,
  \apj, 706, 1364

\bibitem[{{Frayer} {et~al.}(2008){Frayer}, {Koda}, {Pope}, {Huynh}, {Chary},
  {Scott}, {Dickinson}, {Bock}, {Carpenter}, {Hawkins}, {Hodges}, {Lamb},
  {Plambeck}, {Pound}, {Scott}, {Scoville}, \& {Woody}}]{Frayer08}
{Frayer}, D.~T., {Koda}, J., {Pope}, A., {et~al.} 2008, \apjl, 680, L21

\bibitem[{{Genzel} {et~al.}(2010){Genzel}, {Tacconi}, {Gracia-Carpio},
  {Sternberg}, {Cooper}, {Shapiro}, {Bolatto}, {Bouch{\'e}}, {Bournaud},
  {Burkert}, {Combes}, {Comerford}, {Cox}, {Davis}, {Schreiber},
  {Garcia-Burillo}, {Lutz}, {Naab}, {Neri}, {Omont}, {Shapley}, \&
  {Weiner}}]{Genzel10}
{Genzel}, R., {Tacconi}, L.~J., {Gracia-Carpio}, J., {et~al.} 2010, \mnras,
  407, 2091

\bibitem[{{Groves} {et~al.}(2008){Groves}, {Dopita}, {Sutherland}, {Kewley},
  {Fischera}, {Leitherer}, {Brandl}, \& {van Breugel}}]{Groves08}
{Groves}, B., {Dopita}, M.~A., {Sutherland}, R.~S., {et~al.} 2008, \apjs, 176,
  438

\bibitem[{{Groves} {et~al.}(2012){Groves}, {Krause}, {Sandstrom}, {Schmiedeke},
  {Leroy}, {Linz}, {Kapala}, {Rix}, {Schinnerer}, {Tabatabaei}, {Walter}, \&
  {da Cunha}}]{Groves12}
{Groves}, B., {Krause}, O., {Sandstrom}, K., {et~al.} 2012, \mnras, 426, 892

\bibitem[{{Hayward} {et~al.}(2012){Hayward}, {Jonsson}, {Kere{\v s}},
  {Magnelli}, {Hernquist}, \& {Cox}}]{Hayward12}
{Hayward}, C.~C., {Jonsson}, P., {Kere{\v s}}, D., {et~al.} 2012, \mnras, 424,
  951

\bibitem[{{Hayward} {et~al.}(2011){Hayward}, {Kere{\v s}}, {Jonsson},
  {Narayanan}, {Cox}, \& {Hernquist}}]{Hayward11}
{Hayward}, C.~C., {Kere{\v s}}, D., {Jonsson}, P., {et~al.} 2011, \apj, 743,
  159

\bibitem[{{Hayward} \& {Smith}(2014)}]{Hayward15}
{Hayward}, C.~C., \& {Smith}, D.~J.~B. 2014, ArXiv e-prints, arXiv:1409.6332

\bibitem[{{Hayward} {et~al.}(2014{\natexlab{a}}){Hayward}, {Torrey},
  {Springel}, {Hernquist}, \& {Vogelsberger}}]{Hayward14a}
{Hayward}, C.~C., {Torrey}, P., {Springel}, V., {Hernquist}, L., \&
  {Vogelsberger}, M. 2014{\natexlab{a}}, \mnras, 442, 1992

\bibitem[{{Hayward} {et~al.}(2014{\natexlab{b}}){Hayward}, {Lanz}, {Ashby},
  {Fazio}, {Hernquist}, {Rafael Mart{\'{\i}}nez-Galarza}, {Noeske}, {Smith},
  {Wuyts}, \& {Zezas}}]{Hayward14b}
{Hayward}, C.~C., {Lanz}, L., {Ashby}, M.~L.~N., {et~al.} 2014{\natexlab{b}},
  ArXiv e-prints, arXiv:1402.0006

\bibitem[{{Hopkins} {et~al.}(2014){Hopkins}, {Kere{\v s}}, {O{\~n}orbe},
  {Faucher-Gigu{\`e}re}, {Quataert}, {Murray}, \& {Bullock}}]{Hopkins14}
{Hopkins}, P.~F., {Kere{\v s}}, D., {O{\~n}orbe}, J., {et~al.} 2014, \mnras,
  445, 581

\bibitem[{{Hung} {et~al.}(2013){Hung}, {Sanders}, {Casey}, {Lee}, {Barnes},
  {Capak}, {Kartaltepe}, {Koss}, {Larson}, {Le Floc'h}, {Lockhart}, {Man},
  {Mann}, {Riguccini}, {Scoville}, \& {Symeonidis}}]{Hung13}
{Hung}, C.-L., {Sanders}, D.~B., {Casey}, C.~M., {et~al.} 2013, \apj, 778, 129

\bibitem[{{Imanishi} {et~al.}(2007){Imanishi}, {Dudley}, {Maiolino}, {Maloney},
  {Nakagawa}, \& {Risaliti}}]{Imanishi07}
{Imanishi}, M., {Dudley}, C.~C., {Maiolino}, R., {et~al.} 2007, \apjs, 171, 72

\bibitem[{{Jonsson}(2006)}]{Jonsson06}
{Jonsson}, P. 2006, \mnras, 372, 2

\bibitem[{{Jonsson} {et~al.}(2010){Jonsson}, {Groves}, \& {Cox}}]{Jonsson10}
{Jonsson}, P., {Groves}, B.~A., \& {Cox}, T.~J. 2010, \mnras, 403, 17

\bibitem[{{Kartaltepe} {et~al.}(2012){Kartaltepe}, {Dickinson}, {Alexander},
  {Bell}, {Dahlen}, {Elbaz}, {Faber}, {Lotz}, {McIntosh}, {Wiklind}, {Altieri},
  {Aussel}, {Bethermin}, {Bournaud}, {Charmandaris}, {Conselice}, {Cooray},
  {Dannerbauer}, {Dav{\'e}}, {Dunlop}, {Dekel}, {Ferguson}, {Grogin}, {Hwang},
  {Ivison}, {Kocevski}, {Koekemoer}, {Koo}, {Lai}, {Leiton}, {Lucas}, {Lutz},
  {Magdis}, {Magnelli}, {Morrison}, {Mozena}, {Mullaney}, {Newman}, {Pope},
  {Popesso}, {van der Wel}, {Weiner}, \& {Wuyts}}]{Kartaltepe12}
{Kartaltepe}, J.~S., {Dickinson}, M., {Alexander}, D.~M., {et~al.} 2012, \apj,
  757, 23

\bibitem[{{Kelson}(2014)}]{Kelson14}
{Kelson}, D.~D. 2014, ArXiv e-prints, arXiv:1406.5191

\bibitem[{{Kennicutt}(1998)}]{Kennicutt98}
{Kennicutt}, Jr., R.~C. 1998, \araa, 36, 189

\bibitem[{{Krumholz} {et~al.}(2012){Krumholz}, {Dekel}, \&
  {McKee}}]{Krumholz12}
{Krumholz}, M.~R., {Dekel}, A., \& {McKee}, C.~F. 2012, \apj, 745, 69

\bibitem[{{Lanz} {et~al.}(2014){Lanz}, {Hayward}, {Zezas}, {Smith}, {Ashby},
  {Brassington}, {Fazio}, \& {Hernquist}}]{Lanz14}
{Lanz}, L., {Hayward}, C.~C., {Zezas}, A., {et~al.} 2014, \apj, 785, 39

\bibitem[{{Lanz} {et~al.}(2013){Lanz}, {Zezas}, {Brassington}, {Smith},
  {Ashby}, {da Cunha}, {Fazio}, {Hayward}, {Hernquist}, \& {Jonsson}}]{Lanz13}
{Lanz}, L., {Zezas}, A., {Brassington}, N., {et~al.} 2013, \apj, 768, 90

\bibitem[{{Lee} {et~al.}(2013){Lee}, {Sanders}, {Casey}, {Scoville}, {Hung},
  {Le Floc'h}, {Ilbert}, {Aussel}, {Capak}, {Kartaltepe}, {Roseboom},
  {Salvato}, {Aravena}, {Berta}, {Bock}, {Oliver}, {Riguccini}, \&
  {Symeonidis}}]{Lee13}
{Lee}, N., {Sanders}, D.~B., {Casey}, C.~M., {et~al.} 2013, \apj, 778, 131

\bibitem[{{Leitherer} {et~al.}(1999){Leitherer}, {Schaerer}, {Goldader},
  {Delgado}, {Robert}, {Kune}, {de Mello}, {Devost}, \&
  {Heckman}}]{Leitherer99}
{Leitherer}, C., {Schaerer}, D., {Goldader}, J.~D., {et~al.} 1999, \apjs, 123,
  3

\bibitem[{{Lintott} {et~al.}(2008){Lintott}, {Schawinski}, {Slosar}, {Land},
  {Bamford}, {Thomas}, {Raddick}, {Nichol}, {Szalay}, {Andreescu}, {Murray}, \&
  {Vandenberg}}]{Lintott08}
{Lintott}, C.~J., {Schawinski}, K., {Slosar}, A., {et~al.} 2008, \mnras, 389,
  1179

\bibitem[{{Magdis} {et~al.}(2010){Magdis}, {Elbaz}, {Hwang}, {Daddi},
  {Rigopoulou}, {Altieri}, {Andreani}, {Aussel}, {Berta}, {Cava},
  {Bongiovanni}, {Cepa}, {Cimatti}, {Dickinson}, {Dominguez}, {F{\"o}rster
  Schreiber}, {Genzel}, {Huang}, {Lutz}, {Maiolino}, {Magnelli}, {Morrison},
  {Nordon}, {P{\'e}rez Garc{\'{\i}}a}, {Poglitsch}, {Popesso}, {Pozzi},
  {Riguccini}, {Rodighiero}, {Saintonge}, {Santini}, {Sanchez-Portal}, {Shao},
  {Sturm}, {Tacconi}, \& {Valtchanov}}]{Magdis10b}
{Magdis}, G.~E., {Elbaz}, D., {Hwang}, H.~S., {et~al.} 2010, \apjl, 720, L185

\bibitem[{{Magdis} {et~al.}(2012){Magdis}, {Daddi}, {B{\'e}thermin}, {Sargent},
  {Elbaz}, {Pannella}, {Dickinson}, {Dannerbauer}, {da Cunha}, {Walter},
  {Rigopoulou}, {Charmandaris}, {Hwang}, \& {Kartaltepe}}]{Magdis12}
{Magdis}, G.~E., {Daddi}, E., {B{\'e}thermin}, M., {et~al.} 2012, \apj, 760, 6

\bibitem[{{Magdis} {et~al.}(2014){Magdis}, {Rigopoulou}, {Hopwood}, {Huang},
  {Farrah}, {Pearson}, {Alonso-Herrero}, {Bock}, {Clements}, {Cooray},
  {Griffin}, {Oliver}, {Perez Fournon}, {Riechers}, {Swinyard}, {Scott},
  {Thatte}, {Valtchanov}, \& {Vaccari}}]{Magdis14}
{Magdis}, G.~E., {Rigopoulou}, D., {Hopwood}, R., {et~al.} 2014, ArXiv
  e-prints, arXiv:1409.5605

\bibitem[{{Magnelli} {et~al.}(2014){Magnelli}, {Lutz}, {Saintonge}, {Berta},
  {Santini}, {Symeonidis}, {Altieri}, {Andreani}, {Aussel}, {B{\'e}thermin},
  {Bock}, {Bongiovanni}, {Cepa}, {Cimatti}, {Conley}, {Daddi}, {Elbaz},
  {F{\"o}rster Schreiber}, {Genzel}, {Ivison}, {Le Floc'h}, {Magdis},
  {Maiolino}, {Nordon}, {Oliver}, {Page}, {P{\'e}rez Garc{\'{\i}}a},
  {Poglitsch}, {Popesso}, {Pozzi}, {Riguccini}, {Rodighiero}, {Rosario},
  {Roseboom}, {Sanchez-Portal}, {Scott}, {Sturm}, {Tacconi}, {Valtchanov},
  {Wang}, \& {Wuyts}}]{Magnelli14}
{Magnelli}, B., {Lutz}, D., {Saintonge}, A., {et~al.} 2014, \aap, 561, A86

\bibitem[{{Mentuch Cooper} {et~al.}(2012){Mentuch Cooper}, {Wilson}, {Foyle},
  {Bendo}, {Koda}, {Baes}, {Boquien}, {Boselli}, {Ciesla}, {Cooray}, {Eales},
  {Galametz}, {Lebouteiller}, {Parkin}, {Roussel}, {Sauvage}, {Spinoglio}, \&
  {Smith}}]{Cooper12}
{Mentuch Cooper}, E., {Wilson}, C.~D., {Foyle}, K., {et~al.} 2012, \apj, 755,
  165

\bibitem[{{Mitchell} {et~al.}(2014){Mitchell}, {Lacey}, {Cole}, \&
  {Baugh}}]{Mitchell14}
{Mitchell}, P.~D., {Lacey}, C.~G., {Cole}, S., \& {Baugh}, C.~M. 2014, \mnras,
  444, 2637

\bibitem[{{Moshir} {et~al.}(1992){Moshir}, {Kopman}, \& {Conrow}}]{Moshir92}
{Moshir}, M., {Kopman}, G., \& {Conrow}, T.~A.~O. 1992, {IRAS Faint Source
  Survey, Explanatory supplement version 2}

\bibitem[{{Narayanan} {et~al.}(2010{\natexlab{a}}){Narayanan}, {Hayward},
  {Cox}, {Hernquist}, {Jonsson}, {Younger}, \& {Groves}}]{Narayanan10a}
{Narayanan}, D., {Hayward}, C.~C., {Cox}, T.~J., {et~al.} 2010{\natexlab{a}},
  \mnras, 401, 1613

\bibitem[{{Narayanan} {et~al.}(2010{\natexlab{b}}){Narayanan}, {Dey},
  {Hayward}, {Cox}, {Bussmann}, {Brodwin}, {Jonsson}, {Hopkins}, {Groves},
  {Younger}, \& {Hernquist}}]{Narayanan10b}
{Narayanan}, D., {Dey}, A., {Hayward}, C.~C., {et~al.} 2010{\natexlab{b}},
  \mnras, 407, 1701

\bibitem[{{Noeske} {et~al.}(2007){Noeske}, {Weiner}, {Faber}, {Papovich},
  {Koo}, {Somerville}, {Bundy}, {Conselice}, {Newman}, {Schiminovich}, {Le
  Floc'h}, {Coil}, {Rieke}, {Lotz}, {Primack}, {Barmby}, {Cooper}, {Davis},
  {Ellis}, {Fazio}, {Guhathakurta}, {Huang}, {Kassin}, {Martin}, {Phillips},
  {Rich}, {Small}, {Willmer}, \& {Wilson}}]{Noeske07}
{Noeske}, K.~G., {Weiner}, B.~J., {Faber}, S.~M., {et~al.} 2007, \apjl, 660,
  L43

\bibitem[{{Oliver} {et~al.}(2012){Oliver}, {Bock}, {Altieri}, {Amblard},
  {Arumugam}, {Aussel}, {Babbedge}, {Beelen}, {B{\'e}thermin}, {Blain},
  {Boselli}, {Bridge}, {Brisbin}, {Buat}, {Burgarella},
  {Castro-Rodr{\'{\i}}guez}, {Cava}, {Chanial}, {Cirasuolo}, {Clements},
  {Conley}, {Conversi}, {Cooray}, {Dowell}, {Dubois}, {Dwek}, {Dye}, {Eales},
  {Elbaz}, {Farrah}, {Feltre}, {Ferrero}, {Fiolet}, {Fox}, {Franceschini},
  {Gear}, {Giovannoli}, {Glenn}, {Gong}, {Gonz{\'a}lez Solares}, {Griffin},
  {Halpern}, {Harwit}, {Hatziminaoglou}, {Heinis}, {Hurley}, {Hwang}, {Hyde},
  {Ibar}, {Ilbert}, {Isaak}, {Ivison}, {Lagache}, {Le Floc'h}, {Levenson},
  {Faro}, {Lu}, {Madden}, {Maffei}, {Magdis}, {Mainetti}, {Marchetti},
  {Marsden}, {Marshall}, {Mortier}, {Nguyen}, {O'Halloran}, {Omont}, {Page},
  {Panuzzo}, {Papageorgiou}, {Patel}, {Pearson}, {P{\'e}rez-Fournon}, {Pohlen},
  {Rawlings}, {Raymond}, {Rigopoulou}, {Riguccini}, {Rizzo}, {Rodighiero},
  {Roseboom}, {Rowan-Robinson}, {S{\'a}nchez Portal}, {Schulz}, {Scott},
  {Seymour}, {Shupe}, {Smith}, {Stevens}, {Symeonidis}, {Trichas}, {Tugwell},
  {Vaccari}, {Valtchanov}, {Vieira}, {Viero}, {Vigroux}, {Wang}, {Ward},
  {Wardlow}, {Wright}, {Xu}, \& {Zemcov}}]{Oliver12}
{Oliver}, S.~J., {Bock}, J., {Altieri}, B., {et~al.} 2012, \mnras, 424, 1614

\bibitem[{{Ott}(2010)}]{Ott10}
{Ott}, S. 2010, in Astronomical Society of the Pacific Conference Series, Vol.
  434, Astronomical Data Analysis Software and Systems XIX, ed. Y.~{Mizumoto},
  K.-I. {Morita}, \& M.~{Ohishi}, 139

\bibitem[{{Porter} {et~al.}(2014){Porter}, {Somerville}, {Primack}, \&
  {Johansson}}]{Porter14}
{Porter}, L.~A., {Somerville}, R.~S., {Primack}, J.~R., \& {Johansson}, P.~H.
  2014, \mnras, 444, 942

\bibitem[{{R{\'e}my-Ruyer} {et~al.}(2014){R{\'e}my-Ruyer}, {Madden},
  {Galliano}, {Galametz}, {Takeuchi}, {Asano}, {Zhukovska}, {Lebouteiller},
  {Cormier}, {Jones}, {Bocchio}, {Baes}, {Bendo}, {Boquien}, {Boselli},
  {DeLooze}, {Doublier-Pritchard}, {Hughes}, {Karczewski}, \&
  {Spinoglio}}]{Remy_Ruger14}
{R{\'e}my-Ruyer}, A., {Madden}, S.~C., {Galliano}, F., {et~al.} 2014, \aap,
  563, A31

\bibitem[{{Rodighiero} {et~al.}(2011){Rodighiero}, {Daddi}, {Baronchelli},
  {Cimatti}, {Renzini}, {Aussel}, {Popesso}, {Lutz}, {Andreani}, {Berta},
  {Cava}, {Elbaz}, {Feltre}, {Fontana}, {F{\"o}rster Schreiber},
  {Franceschini}, {Genzel}, {Grazian}, {Gruppioni}, {Ilbert}, {Le Floch},
  {Magdis}, {Magliocchetti}, {Magnelli}, {Maiolino}, {McCracken}, {Nordon},
  {Poglitsch}, {Santini}, {Pozzi}, {Riguccini}, {Tacconi}, {Wuyts}, \&
  {Zamorani}}]{Rodighiero11}
{Rodighiero}, G., {Daddi}, E., {Baronchelli}, I., {et~al.} 2011, \apjl, 739,
  L40

\bibitem[{{Saintonge} {et~al.}(2011){Saintonge}, {Kauffmann}, {Wang}, {Kramer},
  {Tacconi}, {Buchbender}, {Catinella}, {Graci{\'a}-Carpio}, {Cortese},
  {Fabello}, {Fu}, {Genzel}, {Giovanelli}, {Guo}, {Haynes}, {Heckman},
  {Krumholz}, {Lemonias}, {Li}, {Moran}, {Rodriguez-Fernandez}, {Schiminovich},
  {Schuster}, \& {Sievers}}]{Saintonge11}
{Saintonge}, A., {Kauffmann}, G., {Wang}, J., {et~al.} 2011, \mnras, 415, 61

\bibitem[{{Sanders} {et~al.}(1991){Sanders}, {Scoville}, \&
  {Soifer}}]{Sanders91}
{Sanders}, D.~B., {Scoville}, N.~Z., \& {Soifer}, B.~T. 1991, \apj, 370, 158

\bibitem[{{Schmidt}(1959)}]{Schmidt59}
{Schmidt}, M. 1959, \apj, 129, 243

\bibitem[{{Scoville} {et~al.}(2015){Scoville}, {Sheth}, {Aussel}, {Vanden
  Bout}, {Capak}, {Bongiorno}, {Casey}, {Murchikova}, {Koda}, {Pope}, {Toft},
  {Ivison}, {Sanders}, {Manohar}, \& {Lee}}]{Scoville15}
{Scoville}, N., {Sheth}, K., {Aussel}, H., {et~al.} 2015, ArXiv e-prints,
  arXiv:1505.02159

\bibitem[{{Snyder} {et~al.}(2013){Snyder}, {Hayward}, {Sajina}, {Jonsson},
  {Cox}, {Hernquist}, {Hopkins}, \& {Yan}}]{Snyder13}
{Snyder}, G.~F., {Hayward}, C.~C., {Sajina}, A., {et~al.} 2013, \apj, 768, 168

\bibitem[{{Sparre} {et~al.}(2014){Sparre}, {Hayward}, {Springel},
  {Vogelsberger}, {Genel}, {Torrey}, {Nelson}, {Sijacki}, \&
  {Hernquist}}]{Sparre14}
{Sparre}, M., {Hayward}, C.~C., {Springel}, V., {et~al.} 2014, ArXiv e-prints,
  arXiv:1409.0009

\bibitem[{{Springel}(2005)}]{Springel05}
{Springel}, V. 2005, \mnras, 364, 1105

\bibitem[{{Springel}(2010)}]{Springel10}
---. 2010, \araa, 48, 391

\bibitem[{{Springel} \& {Hernquist}(2003)}]{Springel03}
{Springel}, V., \& {Hernquist}, L. 2003, \mnras, 339, 289

\bibitem[{{Tacconi} {et~al.}(2013){Tacconi}, {Neri}, {Genzel}, {Combes},
  {Bolatto}, {Cooper}, {Wuyts}, {Bournaud}, {Burkert}, {Comerford}, {Cox},
  {Davis}, {F{\"o}rster Schreiber}, {Garc{\'{\i}}a-Burillo}, {Gracia-Carpio},
  {Lutz}, {Naab}, {Newman}, {Omont}, {Saintonge}, {Shapiro Griffin}, {Shapley},
  {Sternberg}, \& {Weiner}}]{Tacconi13}
{Tacconi}, L.~J., {Neri}, R., {Genzel}, R., {et~al.} 2013, \apj, 768, 74

\bibitem[{{Torrey} {et~al.}(2014){Torrey}, {Vogelsberger}, {Genel}, {Sijacki},
  {Springel}, \& {Hernquist}}]{Torrey14}
{Torrey}, P., {Vogelsberger}, M., {Genel}, S., {et~al.} 2014, \mnras, 438, 1985

\bibitem[{{Wang} {et~al.}(2012){Wang}, {Huang}, {Faber}, {Fang}, {Wuyts},
  {Fazio}, {Yan}, {Dekel}, {Guo}, {Ferguson}, {Grogin}, {Lotz}, {Weiner},
  {McGrath}, {Kocevski}, {Hathi}, {Lucas}, {Koekemoer}, {Kong}, \&
  {Gu}}]{Wang12}
{Wang}, T., {Huang}, J.-S., {Faber}, S.~M., {et~al.} 2012, \apj, 752, 134

\bibitem[{{Weingartner} \& {Draine}(2001)}]{Weingartner01}
{Weingartner}, J.~C., \& {Draine}, B.~T. 2001, \apj, 548, 296

\bibitem[{{Whitaker} {et~al.}(2012){Whitaker}, {van Dokkum}, {Brammer}, \&
  {Franx}}]{Whitaker12}
{Whitaker}, K.~E., {van Dokkum}, P.~G., {Brammer}, G., \& {Franx}, M. 2012,
  \apjl, 754, L29

\bibitem[{{Whitaker} {et~al.}(2014){Whitaker}, {Franx}, {Leja}, {van Dokkum},
  {Henry}, {Skelton}, {Fumagalli}, {Momcheva}, {Brammer}, {Labb{\'e}},
  {Nelson}, \& {Rigby}}]{Whitaker14}
{Whitaker}, K.~E., {Franx}, M., {Leja}, J., {et~al.} 2014, \apj, 795, 104

\bibitem[{{Wuyts} {et~al.}(2011){Wuyts}, {F{\"o}rster Schreiber}, {van der
  Wel}, {Magnelli}, {Guo}, {Genzel}, {Lutz}, {Aussel}, {Barro}, {Berta},
  {Cava}, {Graci{\'a}-Carpio}, {Hathi}, {Huang}, {Kocevski}, {Koekemoer},
  {Lee}, {Le Floc'h}, {McGrath}, {Nordon}, {Popesso}, {Pozzi}, {Riguccini},
  {Rodighiero}, {Saintonge}, \& {Tacconi}}]{Wuyts11}
{Wuyts}, S., {F{\"o}rster Schreiber}, N.~M., {van der Wel}, A., {et~al.} 2011,
  \apj, 742, 96

\end{thebibliography}

\end{document}